\documentclass{aa}  

\usepackage{graphicx}
\usepackage{txfonts}
\usepackage{subcaption}
\usepackage{mhchem}
\usepackage{booktabs}
\usepackage{colortbl}
\usepackage{xcolor}
\usepackage{xspace}
\usepackage{threeparttable}
\usepackage{multirow}
\usepackage{float}

\usepackage{hyperref}
\hypersetup{
    colorlinks=true,
    linkcolor=blue,
    filecolor=blue,      
    urlcolor=blue,
    citecolor=blue
    }
\DeclareRobustCommand{\COdia}{\ce{CO-W2-a}\xspace} 
\DeclareRobustCommand{\COdib}{\ce{CO-W2-b}\xspace}   
\DeclareRobustCommand{\COtra}{\ce{CO-W3-a}\xspace} 
\DeclareRobustCommand{\COtrb}{\ce{CO-W3-b}\xspace} 
\DeclareRobustCommand{\COtrc}{\ce{CO-W3-c}\xspace} 

\newcommand{\kcal}{kcal mol$^{-1}$}

\begin{document} 
\title{CO Adsorption Sites on Interstellar Water Ices\\ Explored with Machine Learning Potentials}

  \subtitle{Binding energy distributions and snowline}

   \author{G. M. Bovolenta
          \inst{1,2}
          \and
          G. Molpeceres
          \inst{3}
          \and
          K. Furuya
          \inst{4}
          \and
          J. K\"astner
          \inst{5}
          \and
          S. Vogt-Geisse
          \inst{1}
          \thanks{Corresponding authors: SVG and GM.}
          }

   \institute{
        Departamento de Físico-Química, Facultad de Ciencias Químicas, Universidad de Concepción, Concepción, Chile 
        \email{stvogtgeisse@qcmmlab.com}
        \and
        Atomistic Simulations, Italian Institute of Technology, 16152 Genova, Italy
        \and
         Departamento de Astrofísica Molecular, Instituto de Física Fundamental (IFF-CSIC), C/ Serrano 121, 28006 Madrid, Spain
        \email{german.molpeceres@iff.csic.es}
        \and
        RIKEN Pioneering Research Institute
        , 2-1 Hirosawa, Wako-shi, Saitama 351-0198, Japan
        \and
        Institute for Theoretical Chemistry, University of Stuttgart, Pfaffenwaldring 55, 70569 Stuttgart, Germany 
             }

   \date{Received \today; accepted \today}

\abstract{}{}{}{}{} 

\abstract
   {Carbon monoxide (CO) is arguably the most important molecule for interstellar organic chemistry. Its binding to amorphous solid water (ASW) ice regulates both diffusion and desorption processes. Accurately characterizing the CO binding energy (BE) is essential for realistic astrochemical modeling.}
   {We aim to derive a statistically robust and physically accurate distribution of CO BEs  on ASW surfaces, and to evaluate its implications for laboratory temperature-programmed desorption experiments and interstellar chemistry, with a focus on protoplanetary disks.}
   {We trained a machine-learned potential (MLP) on 8321 density functional theory  (DFT) energies and gradients of CO interacting with differently-sized water clusters (22-60 water molecules). The DFT method was selected after extensive benchmark. With this potential we built realistic non-porous and porous
  ASW surfaces, and computed a BE distribution. We used symmetry-adapted perturbation theory to rationalize the interaction of CO on the different binding sites. }
   {We find that both ASW morphologies yield similar Gaussian-like BE distributions with mean values near 900 K. However, the nature of the binding interactions is rather different and critically depends on surface roughness and dangling-OH bonds. Simulated  temperature-programmed desorption (TPD) curves reproduce experimental trends across several coverage regimes. From an astrochemical point of view, the application of the  full BE distribution has a dramatic influence on the CO distribution in protoplanetary disks, leading to a broader CO snowline region, improving predictions of CO gas-ice partitioning, and suggesting an equally broader distribution of organics in these objects.}
   {}


   \maketitle
%

\section{Introduction} \label{sec:introduction}

Interstellar surface chemistry plays a pivotal role in the formation of complex organic molecules (COMs) in space \citep{herbst_complex_2009, cuppen_grain_2017}. The icy mantles of dust grains provide a reactive environment where atoms and molecules can accrete, diffuse, and interact, facilitating the synthesis of new species and driving molecular complexity in the interstellar medium (ISM). Within these surface-driven networks, certain molecules act as key nodes, either as chemically stable reservoirs (e.g., \ce{CH4}, \ce{CO2}, \ce{N2}) or as reactive intermediates that enable further chemical evolution. Among the latter, carbon monoxide (CO) is particularly important \citep{Watanabe2002, fuchs_hydrogenation_2009, fedoseev_experimental_2015, fedoseev_cosmic_2017, simons_formation_2020}, serving as the principal precursor to all oxygen-bearing COMs. 

Accurately modeling surface processes requires reliable input parameters. In addition to the intrinsic reactivity of the species involved, one of the most critical parameters is the binding energy (BE), which quantifies the strength of interaction between adsorbed molecules and the ice surface. This value directly influences the desorption rate and indirectly the surface diffusion and, consequently, reaction rates; for a comprehensive overview, see \citet{cuppen_grain_2017}. 

BEs on ice surfaces are typically obtained through temperature-programmed desorption (TPD) experiments. During the last two decades, values have been measured for a plethora of astrophysically relevant small species \citep{collings_laboratory_2004, amiaud_interaction_2006,noble_1_2012,smith_desorption_2016,he_binding_2016,nguyen_segregation_2018,chaabouni_thermal_2018,nguyen_experimental_2020}. 
However, differences in substrate preparation, surface coverage and deposition methods limit comparability among experimental results. 

Recent efforts have focused on using \textit{ab initio} quantum chemistry 
methods to compute binding energies across a variety of interstellar substrates \citep{wakelam_binding_2017, das_approach_2018, molpeceres_adsorption_2020, ferrero_binding_2020, bovolenta_high_2020, 
duflot_theoretical_2021,sameera_ch3o_2021, bovolenta_binding_2022,
tinacci_theoretical_2022, perrero_binding_2022-1, molpeceres_enhanced_2024}. A key advancement in our understanding of interstellar surface chemistry is the recognition of binding site heterogeneity. Rather than a single fixed value, BEs are now understood to follow distributions, especially on amorphous solid water (ASW), where the disordered arrangement of water (\ce{H2O}) dipoles gives rise to a broad spectrum of adsorption environments.

Among the various approaches to simulate adsorption sites on interstellar ASW,  one of the most widely used and general methods involves modeling the ice as a finite cluster composed of a handful of water molecules \citep{wakelam_binding_2017, das_approach_2018, bovolenta_high_2020, bovolenta_binding_2022, tinacci_theoretical_2022,  molpeceres_enhanced_2024, bovolenta_-depth_2024, bovolenta_methyl_2025}. The number of molecules needed to adequately capture the BE, and especially the distribution of BEs, depends strongly on the nature of the adsorbate - surface interaction. For species less volatile than water ice, precise BE calculations are often of limited practical relevance, since such molecules tend neither to diffuse nor desorb significantly under typical interstellar conditions, at least when adsorbed on pure \ce{H2O} substrates.\footnote{The heterogeneous nature of interstellar grain surfaces in astrochemical models is an active area of research; see, e.g., \citet{molpeceres_enhanced_2024, kalvans_multi-grain_2024}.} By contrast, the mobility of more volatile species is critical in cold interstellar environments. A prominent example is CO, the focus of this study. Other key mobile species include atomic H, O, and N, as well as molecular \ce{H2}. Heavier radicals have also been recognized as important agents in surface chemistry at low temperatures \citep{furuya_diffusion_2022}. For such species, cluster models may fall short in capturing the full diversity of binding sites, especially compared to adsorbates with strong dipole moments that can preferentially orient along the polar ASW network, leading to more predictable binding motifs.

An important physical characteristic of ice mantles is their porosity, though its degree remains under debate. Laboratory measurements \citep{dohnalek_deposition_2003,bossa_porosity_2014}  and theoretical studies \citep{cuppen_simulation_2007,clements_kinetic_2018} suggest porosity formation in cold interstellar environments, particularly at the surface/subsurface level of the ASW. In contrast, processes such as UV radiation, exothermic chemical reactions, and ion impacts are expected to have a compacting effect,  as suggested also by 
the apparent absence of O–H dangling bond features in astronomical spectra \citep{keane_bands_2001,hama_surface_2013}.   
This view, however, has been challenged a tentative detection of the O–H dangling feature \citep{mcclure_ice_2023}, and definite confirmation \citep{noble_detection_2024}, using  
the James Webb Space Telescope.
Employing a larger periodic model might be more appropriate for accurately simulating ASW mantles than cluster or small slab models, 
as it enables the representation 
of structural features such as nanoscale pores.
Dealing with large-size systems, however, is challenging, due to the prohibitive computational cost.
In recent years, machine learning potentials (MLP) have emerged as a promising strategy to combine the accuracy of \textit{ab initio} methods with the computational efficiency of classical force fields. 
These potentials are trained to reproduce energies and forces obtained from a extensive set quantum mechanical calculations and, once optimized on small systems, can be applied to the simulation of larger ones. 
Despite their advantages, training reliable MLP is a complex task, that requires to carefully assembling a training set that represents all relevant configurations of the system under study. 
Based on this approach, in the astrophysical context, tailored MLPs have been used  to model the interaction and 
binding energies of species on ASW, 
as well as reaction dynamics on ASW surfaces \citep{molpeceres_neural-network_2020,zaverkin_gaussian_2020,molpeceres_reaction_2023,bovolenta_-depth_2024,postulka_diffusive_2025}.         

The aim of this work is to provide an accurate distribution of the BEs of CO on ASW, using a large-scale ice model (1500 atoms; see Section~\ref{sec:methods}) and a high-quality interaction potential. To meet these requirements, we develop and train an interatomic MLP based on density functional theory (DFT) calculations. These DFT data are benchmarked against highly accurate reference methods to ensure that the key intermolecular interactions are reliably captured. As we demonstrate below, this approach enables the sampling of a wide range of CO adsorption sites on extended systems leading to a BE distribution.
Overall, our work aims to establish a new standard for studying CO chemistry on polar ices by profiting from state-of-the-art machine learning techniques. 
In addition, we explore the astrochemical implications of our results by using the derived BE distributions to simulate CO gas-ice partitioning in protoplanetary disks.

After a description of the methods, the analysis tools  and the computational protocol used in this work     (Section \ref{sec:methods}), we present our results  (Section \ref{sec:results}) in terms of the fundamental interaction 
and its statistical nature
of CO binding on ice surface models ranging from small \ce{H2O} clusters to porous and non-porous ASW.    
BEs results and simulated TPD curves are compared to experimental and theoretical findings and we also discuss their astrophysical implications on the predicted protoplanetary disks' CO snowlines  (Section \ref{sec:discussion}) and other interstellar regions. Finally, we present our conclusions  (Section \ref{sec:discussion:conclusions}).

\section{Methodology} \label{sec:methods}
\subsection{Benchmark and model chemistry selection}\label{sec:MIPM} 

In order to select the most appropriate DFT model
chemistry for the training of the MLP, we performed a
comprehensive benchmark to evaluate the performance 
of a plethora of DFT methods in computing both gradients and energy compared to a high-level
coupled cluster reference.
The detailed results of the  benchmark study can be found in Appendix \ref{sec:ap_DFTbench}. 

The electronic binding energy of the molecule adsorbed on different-sized \ce{H2O} ($W$) ice surfaces 
was calculated using the following relationship:
\begin{equation}
    \Delta E_\mathrm{b} = E_{\mathrm{CO-W}_n} - (E_\mathrm{CO} + E_{\mathrm{W}_n}),
\end{equation}
where $E_\mathrm{CO-W}$ is the total energy of the complex formed by the molecule and the surface in an equilibrium geometry; $E_\mathrm{CO}$ and $E_{\mathrm{W}_n}$ represent the energies of the isolated molecule and a 
water surface model composed of $n$ water molecules W$_n$. 
For the model systems we optimized the W$_n$+CO ($n = 2,3$) complex at the DF-CCSD(T)-F12/cc-pVDZ-F12\citep{gyorffy_analytical_2018, sylvetsky_aug-cc-pvnz-f12_2017} 
level of theory using the \textsc{Molpro} \citep{werner_molpro_2012, werner_molpro_2020} atomic gradients in conjunction with the \textsc{GeomeTRIC}\citep{wang_leepinggeometric_2023} optimizer.
We computed high level \textit{ab initio} energies using the CCSD(T)\citep{raghavachari_fifth-order_1989, bozkaya_analytic_2017} method with basis set extrapolation (CBS) extrapolation \citep{helgaker_basis-set_1997-2} of aug-cc-pVXZ (X=D,T,Q) type basis sets \citep{DunningT.H.2001}
The extrapolation formula is outlined in Appendix \ref{sec:ap_DFTbench}.

The overall best-performing 
DFT exchange and correlation functional is MPWB1K-D3BJ \citep{Zhao2005, grimme_consistent_2010, grimme_effect_2011} paired with a def2-TZVP basis \citep{weigend_balanced_2005}. The benchmarks 
were conducted utilizing the workflows implemented in 
the Binding Energy Evaluation Platform (BEEP) \citep{bovolenta_binding_2022}. Basis set superposition 
error (BSSE) was accounted for using the counterpoise method \citep{boys_calculation_1970}. The total method can be therefore denoted as MPWB1K-D3BJ/def2-TZVP.
Finally, using high-level DF-CCSD(T)-F12/cc-pVTZ-F12 geometries of model systems, 
we computed the Hessian matrix to obtain an accurate Zero-point vibrational (ZPVE) correction. 
We also included anharmonic vibrational effects to the ZPVE at the CCSD(T)/cc-pVDZ level of theory and derived a average scaling factor ($f_\mathrm{ZPVE}$) of 0.677, that we applied to all the BEs reported in this work: 
\begin{equation}
    BE = -(\Delta E_\mathrm{b}\,  f_\mathrm{ZPVE})
\end{equation}

For more  details about this procedure and results see Appendix \ref{sec:ap_zpve}.

\subsection{Interaction energy decomposition}\label{sec:ie}

The BE for a species \textit{i}, \textit{BE(i)}, can be decomposed into two components: 
\begin{equation}\label{eq:BE_IE}
BE(i) = IE(i) - DE(i).
    \end{equation}
In the additive relation above, the deformation energy, $DE(i)$, reflects the influence of the substrate morphology and the structural rearrangements required to accommodate an adsorbate. In contrast, the interaction energy, $IE(i)$, is determined by the physicochemical nature of the forces responsible for binding the molecule to the surface. This interaction can be further decomposed based on the dominant intermolecular forces involved. Such a decomposition is valuable for rationalizing the underlying nature of the adsorbate–substrate interaction and contributes to building a general framework for understanding gas–grain  
chemistry in cold interstellar environments.
    
In this work, we computed the interaction energy using zeroth-order Symmetry Adapted
Perturbation Theory methods SAPT0-D3MBJ and SAPT2+ analysis 
\citep{jeziorski_perturbation_1994,szalewicz_symmetry-adapted_2012, schriber_optimized_2021}, 
together with a jun-cc-pVDZ \citep{papajak_perspectives_2011} basis set. This 
basis set has been found to perform best with SAPT0 and SAPT0-D3MBJ levels of 
theory \citep{parker_levels_2014, schriber_optimized_2021}. SAPT0, SAPT0-D3MBJ and
SAPT2+ calculations are performed on cluster models of different 
geometry (W$_n$) and on the binding sites cut-outs comprising 28 water molecules. 
In a CO-bound configuration, in accordance to equation \ref{eq:BE_IE}, 
the interaction energy is a positive defined quantity:
\begin{equation}
IE(\mathrm{CO-W}_n) = - IE(\ce{CO} \dotsb n\mathrm{W})
\label{eq:ie}
\end{equation}
The SAPT energy decomposition allows to partition $IE(\mathrm{CO-W}_n)$ into electrostatic ($E_\mathrm{elest}$), induction ($E_\mathrm{ind}$), dispersion ($E_\mathrm{disp}$) and exchange-repulsion ($E_\mathrm{exch}$). Within this decomposition analysis, the total interaction
energy is given by:

\begin{equation}\label{eq:IE_terms}
IE(\mathrm{CO-W}_n) = E_\mathrm{elst} + E_\mathrm{exch} + E_\mathrm{ind} + E_\mathrm{disp} 
 \end{equation} 
To quantify the relative importance of dispersion, we defined a dispersion 
factor ($F_\mathrm{disp}$) as the ratio of the dispersion energy to 
the total attractive interaction energy  $IE(\mathrm{CO-W}_n)$:

\begin{equation}\label{eq:F_disp}
 F_\mathrm{disp} = \frac{E_\mathrm{disp}}{(E_\mathrm{elst} + E_\mathrm{ind} + E_\mathrm{disp})}.
 \end{equation}

Based on \( F_{\text{disp}} \) values, the structures can be divided into different classes according to which contribution plays the most dominant role in binding CO to the ASW surface. Based on the quartiles of the distribution, the dataset was divided into three classes:  

\begin{itemize}
    \item \makebox[6cm][l]{Primarily Electrostatic (Elst-Class):} \( F_{\text{disp}} < 0.4 \)
    \item \makebox[6cm][l]{Primarily Dispersion (Disp-Class):} \( F_{\text{disp}} > 0.6 \)
    \item \makebox[6cm][l]{Electrostatic and Dispersion (ED-Class):} \( 0.4 < F_{\text{disp}} \leq 0.6 \)
\end{itemize}

\subsection{Machine-learned interatomic potential model}\label{sec:MIPM} 
We used a MLP 
to mimic a realistic ASW interstellar ice-surface (and partially the mantle) representation, while 
maintaining the accuracy that our chosen DFT model chemistry provides.
We trained and ad hoc MLP  tailored to reproduce the
CO-ASW situation, using the Gaussian-Moment Neural Network 
(GMNN) method (\citet{zaverkin_gaussian_2020,zaverkin_fast_2021}). 
To generate the training set configurations, we 
performed molecular dynamics (MD) and metadynamics simulations of water clusters and 
CO-W$_n$
 clusters at different temperatures and then labeled using DFT energy and forces for a subset of configurations extracted from 
the MD trajectories. Additionally we enlarged this training set using the Query-by-Committee (QBC; \cite{seung_query_1992}) approach using an ensemble of pre-trained MLPs. We label this training subset ``Active learning''.  Appendix \ref{sec:ap_trainingset} collates the 
details concerning training set composition and propagation methods.
Training set labeling was done using the 
MPWB1K-D3BJ-gCP/def2-TZVP method selected in our benchmark, incorporating BSSE corrections through
the geometric counterpoise (gCP) \citep{kruse_geometrical_2012} correction, which is an approximation 
of the previously used Boys and Bernardi protocol \cite{boys_calculation_1970}, that we cannot fully incorporate within our training set due to computational constraints. All training set
computations were done using the \textsc{Orca 5.0.4} software \citep{neese_orca_2020}.
The neural network hyper-parameters used in the training of the MLP reflect those shown in the original work \citep{zaverkin_fast_2021}. The MLP was trained for 1000 training epochs. 

We used the QBC approach also in our production simulations, simultaneously training an
ensemble of three MLP models using the same training data but 
with different randomly initialized  seeds. A detailed evaluation 
of the accuracy of the ensemble is provided in Appendix \ref{sec:ap_MLP}.

\subsection{ASW periodic surfaces modeling}\label{sec:surface_model}

We used the MLP to built a set of periodic ice models composed of 500 water molecules each. We chose a ice model size that allows to 
account for the diverse morphological and energetic characteristics of the binding sites.
We simulated an initial tridimensional cell of volume (\textit{X} $\times$ \textit{X} $\times$ \textit{X}/2~\AA$^{3}$)
where $X$ depends on the desired ice density $\rho$. The first surface type with a density of $\rho=0.9982$ g $\mathrm{cm^{-3}}$ 
(dimensions 31.6 $\times$ 31.6 $\times$ 15.8  ~\AA$^{3}$), corresponds to a non-porous ASW surface (\textit{np}ASW). This 
density value is between the experimentally reported density 
values of high and low-density ASW \citep{mariedahl_x-ray_2018}.
In order to introduce porosity into the ice matrix, we 
lowered the density of the \textit{np}ASW to a value of 0.60 g $\mathrm{cm^{-3}}$
, which yields the second porous surface type labeled \textit{p}ASW (dimensions  36.8 $\times$ 36.8 $\times$ 18.4 ~\AA$^{3}$).
After an initial minimization, the systems were equilibrated in the canonical ensemble (NVT) for 100 ps at 300 K. 
We extracted 5 structures from the resulting trajectories, that is $\tau_{\rm correlation}$ $\simeq$ 20 ps),
and performed a temperature annealing of 10 ps to reach interstellar conditions ($\sim 10$ K).
We applied periodic boundary conditions in 2 directions (X, Y) profiting from a transfer from local to periodic models \citep{Zaverkin2022}. We also fixed the bottom end in the Z direction (approximately one third of the water molecules), to reproduce the bulk of the ice models. The applied constraint 
ensures that the specific structural diversity  of the surfaces is preserved. 
The frozen bulk condition is maintained in all the following calculations.
An altitude map of a representative surface is reported in Fig. \ref{fig:ice_asw} (details on Tri-Surface plots are given in Appendix \ref{sec:ap_asw}). 
All the simulations use a Langevin thermostat 
with a time step of 0.5 fs and a friction coefficient of 0.02 fs$^{-1}$.   
To run MD simulations with the MLP, the GMNN program was interfaced to the ASE package \citep{hjorth_larsen_atomic_2017}.

\subsection{Binding site sampling and optimization, binding energy distributions}\label{sec:methods_be}
After preparing  \textit{np}ASW and \textit{p}ASW models, we proceeded with  
sampling both surfaces with the CO molecule.
The sampling algorithm is composed of 5 steps. 

\begin{enumerate}
\item Generate a grid of equispaced points as initial guesses for the CO positions on the ice surface. Each new point is located 3\text{\AA} from its nearest neighbor on the grid. We introduced noise on the grid points by relaxing the grid coordinate by $\pm 1/4$ of the step size in both X and Y directions.
\item  Place the molecule at a grid point and adjust its position in order for the center of mass of the molecule to be placed at a specific distance 
from the nearest neighboring surface atom. We performed three rounds of sampling with initial distances of 2.50, 2.75 and 3.00 \text{\AA} from the surface grid point. The initial CO orientation is randomized through a series of random Euler rotations.
\item  Fully optimize the initial structure  using the  MPL. 
We noted that the final number of binding sites vary for each system, because the cell volume is different based on the cell density. 
\item  Filtered the minima belonging to an initial surface model (resulted from the three sampling rounds) grouping the structures based on a similarity threshold of root-mean-square deviation of atomic
positions (RMSD) $\leq$ 0.07 \text{\AA}).
\end{enumerate}

The resulting BE distribution (BE$_d$) underlying points have been  classified  into five groups, based on how far they deviate from the average BE - avg(BE$_d$) -  using quarter (25\%) intervals:  
\begin{itemize}
    \item \makebox[2.5cm][l]{Very Low (VL):} \( \text{BE} < \text{avg(BE}_d) \times 0.5 \)
    \item \makebox[2.5cm][l]{Low:} \( \text{avg(BE}_d) \times 0.5 < \text{BE} < \text{avg(BE}_d) \times 0.75 \)
    \item \makebox[2.5cm][l]{Medium:} \( \text{avg(BE}_d) \times 0.75 < \text{BE} < \text{avg(BE}_d) \times 1.25 \)
    \item \makebox[2.5cm][l]{High:} \( \text{avg(BE}_d) \times 1.25 < \text{BE} < \text{avg(BE}_d) \times 1.5 \)
    \item \makebox[2.5cm][l]{Very High (VH):} \( \text{BE} > \text{avg(BE}_d) \times 1.5 \)
\end{itemize}

Finally, we used a bootstrap method to obtain the mean ($\mu$ and standard deviation ($\sigma$) of the BE$_d$, taking into account the uncertainties derived from the MLP BE computations 
(Appendix \ref{sec:ap_bootstrap} shows the details of this protocol).

\section{Results} \label{sec:results}

\subsection{Fundamental nature of the CO-ASW interaction}

\subsubsection{Binding modes and binding energies in \ce{CO-W_{2-3}} model systems} 

\begin{figure*}
    \centering
    \includegraphics[width=\linewidth]{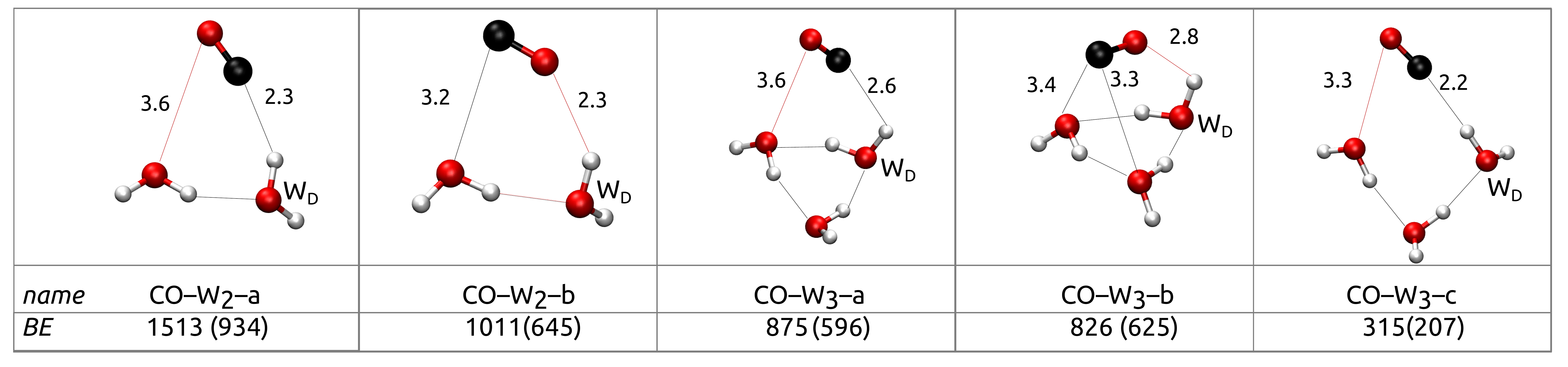} \\
    \caption{\ce{CO-W_{2-3}} model systems that we used for the DFT energy and geometry benchmark. Rows report structure names and binding energies (BE). The numbers in parenthesis indicate the ZPVE corrected BE value. The water molecule acting as H-bond donor to CO is labeled $\mathrm{W_D}$. Characteristic distances are in \text{\AA}; energy values are in K. 
    Color code is black for C, red for O, white for H.}
    \label{fig:be_model}
\end{figure*}

Carbon monoxide has a weak dipole moment of 0.112 Debye \citep{Muenter1975}, with a small
partial negative charge on the carbon end.  This slightly predisposes CO to bind to a polar water ice surface in a low temperature environment.  In order to shed light on this interaction and identify different binding modes, we probed the CO interaction on small water clusters (water dimer and trimer, \ce{W_{2-3}}).  The different binding modes are shown in Fig. \ref{fig:be_model} together with the ZPVE 
corrected BEs. The BEs of all the identified
binding modes lie between 300 K and 1500 K with CO establishing weak hydrogen bonds (H-bonds) 
to the water molecules both through the carbon- and oxygen-end. However, 
the binding mode  with the highest BE  corresponds to the dimer 
structure in which an interaction is established through the carbon end of CO that harbors 
the partial negative charge (\ce{CO-W2-a}).  Interestingly, the equivalent binding mode
in the trimer structure (\ce{CO-W3-a}) shows a decrease in the
BE of 638 K. The addition of a third water molecule results in higher 
rigidity by creating a H-bond cycle; as a consequence, the CO
molecule's interaction with the pair of water molecules becomes less
energetically favorable. 
To gain further insights on the drivers of the CO binding, we decompose the contributions to the BE in the appendix  (Table \ref{tab:IE_DE_BE}). The structure with the lowest BE (\ce{CO-W3-c}) uniquely
exhibits a notable deformation energy, as it disrupts the
H-bond cycle of the water trimer. However, this structure still has 
a positive BE, due to a very favorable interaction energy (the highest 
among the five binding sites) as it takes part in a cyclic structure. 
This is noteworthy because these kinds of H-bond network imperfections
can occur naturally in ASW owing to adjacent water molecules, thereby creating a high BE  environment for CO.

\subsubsection{SAPT analysis of the CO binding}

Table \ref{tab:sapt_model} reports the dispersion 
factor ($F_\mathrm{disp}$, Eq. \ref{eq:F_disp}) that we calculated for the model systems using SAPT at different
levels of accuracy, as an indication of the dispersion character of the binding, while the 
different contributions are reported in Appendix Table \ref{tab:sapt_ref}. The dispersion 
factor is small for the model systems structures ranging between 0.21-0.53, meaning that in 
the model systems CO binds to the small water clusters predominantly through electrostatic and induction 
interactions. Only in \COtrb does the dispersion interaction represents more than 50\% of the attractive
contributions of the interaction energy. The electrostatic and induction attractive 
contributions are larger when CO is interacting with a dangling-OH bond (water H-atoms that are not engaged in any H-bond with other water molecules) via the C-extremity,  
as in the water dimer \ce{CO-W2-a} and trimer \ce{CO-W3-a,c}. These structures also display the 
highest net interaction energy. However, the electrostatic contribution in \ce{CO-W2-a} and
\ce{CO-W3-c}, is more than twice that of \ce{CO-W3-a}. Such enhanced interaction in those structures 
- in which CO is de facto interacting only with a water dimer - is attributable to the fact 
that the H-bond established to CO is the sole donor interaction for that water molecule (namely $\mathrm{W_D}$ in Fig. \ref{fig:be_model}).  
On the other hand, in \ce{CO-W3-a}, 
where a H-bond cycle is established in the water trimer, $\mathrm{W_D}$ is also acting as H-bond donor
to a second water molecule.  Such exclusive donor interaction, established in \ce{CO-W2-a} and \ce{CO-W3-c}, 
is reflected in the closer proximity of CO molecule to $\mathrm{W_D}$, 
and renders $\mathrm{W_D}$ more electron deficient, intensifying the electrostatic coupling with the 
negatively charged carbon-end.
Analysis of the dispersion interaction reveals that the contribution is significantly
lower than the electrostatic/induction one. Nevertheless, for \COtrb 
the balance between dispersion and electrostatic/induction interactions is slightly shifted 
towards dispersion, resulting in a dispersion factor of 0.53. This is consistent with the 
binding mode in \COtrb in which CO is positioned within the triangle formed by the H-bond 
network of the water molecules in the trimer cluster. 
However, due to the closer proximity of the CO molecule to the closest water trimer in \COtrb,
the exchange repulsion is quite significant and higher than in \ce{CO-W3-a} (See Table \ref{tab:sapt_model}). 
Finally, a comparison of the $F_\mathrm{disp}$ values obtained using SAPT/jun-cc-pVDZ and SAPT0-D3MBJ/aug-cc-pVDZ against 
the more accurate SAPT2+/aug-cc-PVDZ reveals that SAPT0-D3MBJ  yields a $F_\mathrm{disp}$ that aligns significantly better with 
the reference value, as opposed to SAPT0. 

\begin{table}[h]
\begin{center}
\caption{Interaction energies (IE) obtained 
at the different symmetry adapted perturbation (SAPT) levels of theory. 
}  
\label{tab:sapt_model}
\begin{tabular*}{0.49\textwidth}{@{\extracolsep{\fill}}lccll}

\toprule
System & IE$_{CCSD(T)}$  &   Method &    IE$_{SAPT}$ & $F_\mathrm{disp}$  \\
\bottomrule
\multirow{3}{*}{\COdia} &  &  SAPT0 & 1152 & 0.21  \\
                &1647 &        SAPT0-D3MBJ& 1656 & 0.30 \\
                  &         & SAPT2+ &  1651  & 0.27 \\
                      \hline  
\multirow{3}{*}{\COdib}  & &  SAPT0 & 1317 &  0.22 \\
               & 1097        & SAPT0-D3MBJ  & 1692 & 0.32  \\
               & &           SAPT2+ &   840 & 0.39 \\
               \hline
\multirow{3}{*}{\COtra} &  & SAPT0 & 656 & 0.30  \\
               & 960           & SAPT0-D3MBJ &  978 &  0.41  \\
                     &   &     SAPT2+ & 985 &  0.38  \\
                     \hline
\multirow{3}{*}{\COtrb} & & SAPT0 & 829 &  0.34\\
               & 880     & SAPT0-D3MBJ & 1238 & 0.47 \\
                &       &    SAPT2+ &    726 &  0.53 \\
                \hline
\multirow{3}{*}{\COtrc} &  & SAPT0 & 1431 & 0.20  \\
                & 1907       & SAPT0-D3MBJ & 1981 & 0.28 \\
                  & &        SAPT2+ &  1852 &  0.26  \\
 \bottomrule
\end{tabular*}
\tablefoot{
The last column shows the dispersion factor $F_\mathrm{disp}$ defined in Eq. \ref{eq:F_disp}. The basis sets used are 
jun-cc-pVDZ for SAPT0 and SAPT0-D3BJ and aug-cc-pVDZ for SAPT2+. The IE at the CCSD(T)/CBS 
level of theory is shown as reference.
}
\end{center}
\end{table}

\subsection{ASW surface characterization}  \label{sec:results:surface}
 We modeled two sets of ASW surfaces following the procedure in Sec. \ref{sec:surface_model}. The first set, that we labeled non-porous (\textit{np}) ASW is composed of smooth surfaces, while the second set, porous (\textit{p}) ASW, presents     
uneven surfaces, with presence of protrusions and cavities. 
In order to characterize and visualize the ice models, 
we have reconstructed the interface using a Tri-Surface interpolation method (see Appendix \ref{sec:ap_asw}). This allows us to identify the surface atoms and to quantify properties such as the areal average roughness(\textit{Sa}, eq. \ref{eq:Ra}), 
and the mean roughness depth (\textit{Rz}, eq. \ref{eq:rz}). The \textit{Sa},   
estimated as the sum of the deviations from the mean plane of the surfaces, entails a value of 3.0 \text{\AA} for the \textit{np}ASW model, being larger for the \textit{p}ASW model, reaching 3.8 \text{\AA}. 
\textit{Rz}, the average of the five largest distances in altitude, yields a value of 15.8 \text{\AA} for \textit{np}ASW and 18.4 \text{\AA} for \textit{p}ASW. These numbers confirm that the \textit{p}ASW model presents larger surface area and deep hollow regions resembling nanopores.    
Such nanopores are spherical-like open-cavities with the pores' width ranging from 2 up to 15 \text{\AA} and the ratio between the height of the cap (\textit{L}) and the radius of the sphere cavity (\textit{Rs}) is \textit{L/Rs} $<$ 1. 
The degree of roughness is relevant since it  has been suggested that 
 for high degrees of roughness (i.e., for
almost closed cavities \textit{L/Rs} $\sim$ 2), molecules desorbing from the grains have high probability of
colliding with the pores' walls, resulting in re-adsorption \citep{maggiolo_effect_2019}.  
The magnitude of the value of \textit{Rz} for \textit{p}ASW models, suggests the definition of valleys and peaks regions, 
namely the portions of the surface that differ the most in terms of altitude and offer the most variegate scenario in terms of binding sites for the adsorbate.         
Another factor at play for the CO adsorption is the presence of dangling-OH bond. 
Typically, such dangling-OH bonds are pointing upward with respect to the surface, and are pivotal in determining the binding site characteristics \citep{al-halabi_adsorption_2004, nagasawa_absolute_2021, bovolenta_-depth_2024}.  
Estimation of the percentage dangling-OH bonds ($\%d(OH)$) reveals notable difference between the models: $\%d(OH)$ is  
6.9 and 10.8 in favor of the \textit{p}ASW.  This is due to the fact that the \textit{np}ASW model is more compact and a larger number of H-bond interactions are established within the ice, decreasing the amount of dangling-OH bonds. Appendix Fig. \ref{fig:dOH_p} shows the location of the dangling-OH bonds in the different models. It is worth noticing that dangling-OH bonds are present in the nanopores inner surface, as well.      

\subsection{Statistical nature of CO interaction with water}

\begin{figure}
    \centering
    \includegraphics[width=1.\linewidth]{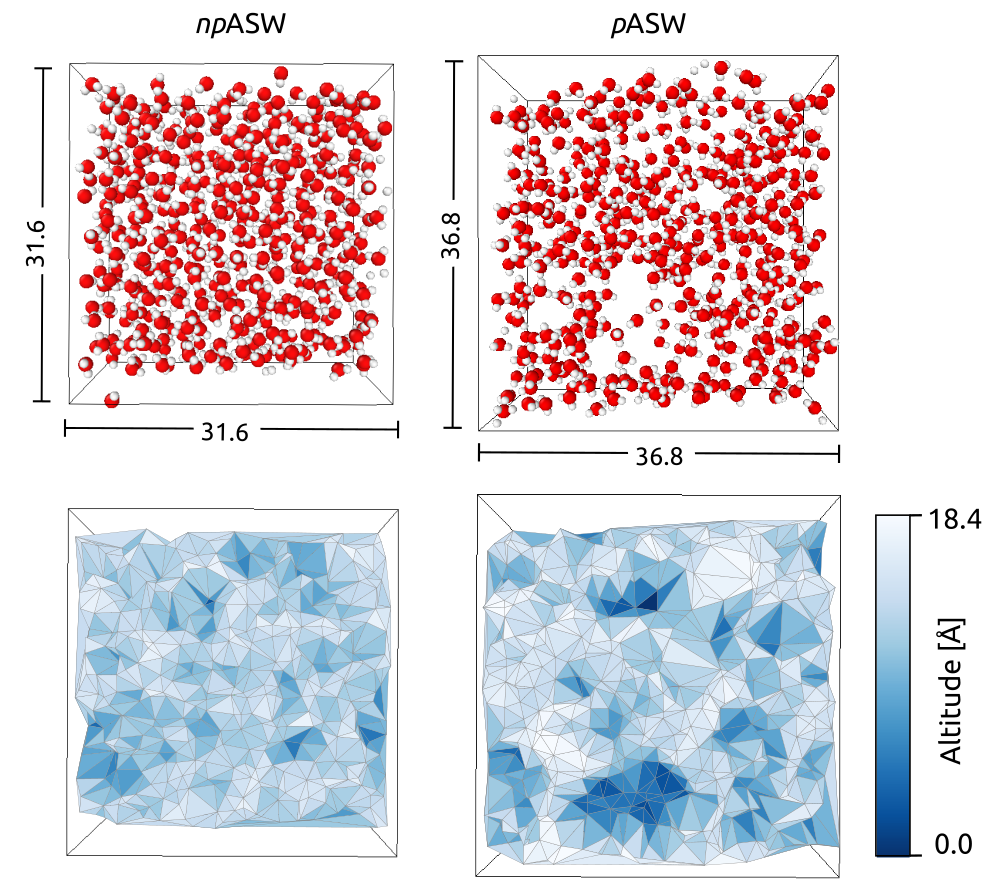} \\
    \caption{Upper panel: top view of one of the non-porous (\textit{np}ASW)  (left) and porous (\textit{p}ASW) ice models. Each periodic surface contains 500 water molecules.  Lower panel: altitude plot relative to the same two structures (see Appendix \ref{sec:ap_asw}). Periodic cell dimensions are included. Distances in \text{\AA}.}
    \label{fig:ice_asw}
\end{figure}

\subsubsection{Binding energy distributions}  \label{sec:results:be}
\begin{figure}
    \centering
    \includegraphics[width=\linewidth]{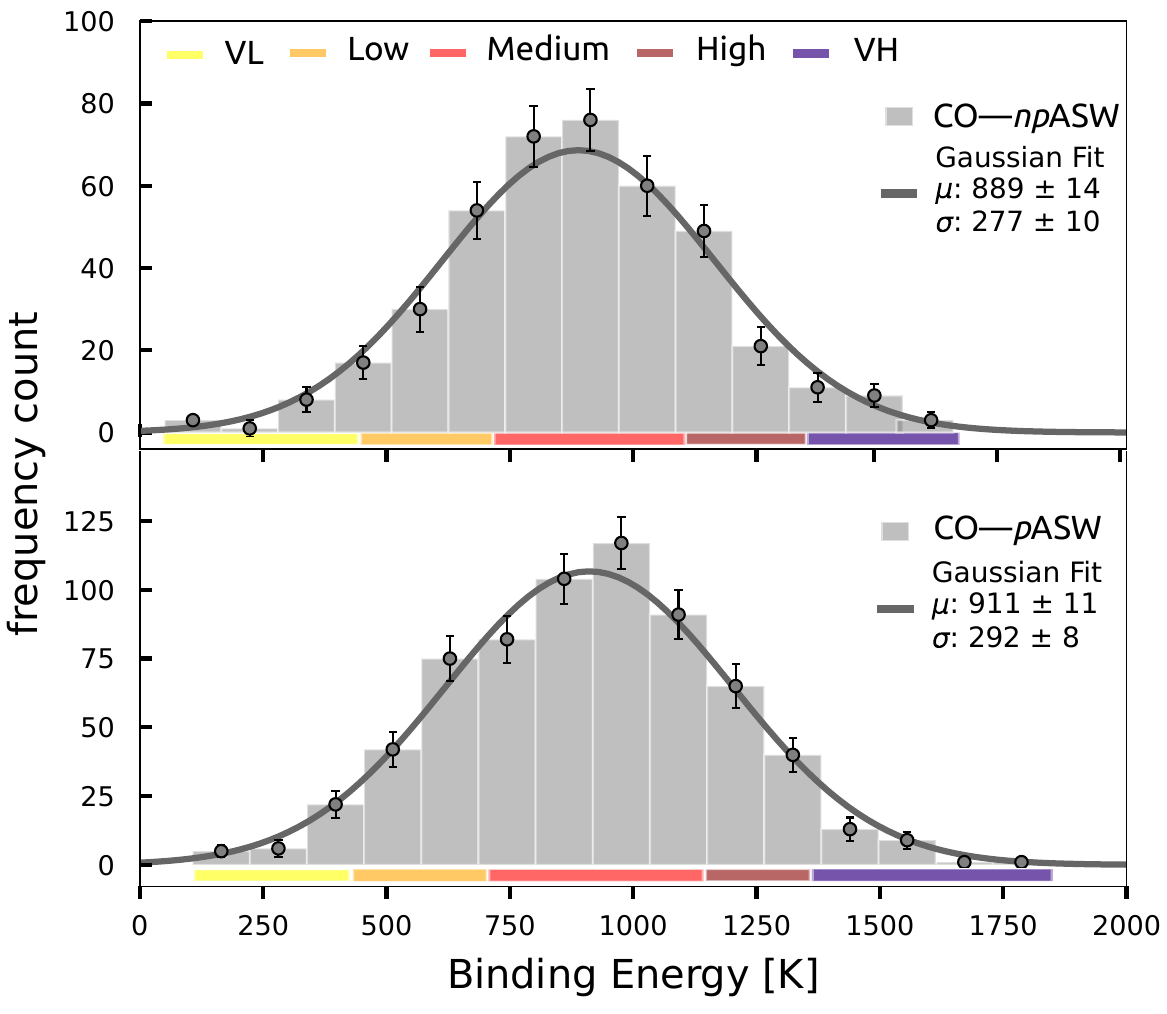} \\
\caption{Histograms of the ZPVE corrected BEs obtained for CO adsorbed on five \textit{np}ASW (upper panel) and five \textit{p}ASW surfaces (lower panel). The color‐coded bars below each panel indicate the BE groups (VL: Very Low, Low, Medium, High, and VH: Very High). The Gaussian fits were obtained using 
the bootstrap procedure detailed in Appendix \ref{sec:ap_bootstrap}, that propagates the individual uncertainties from the neural-network models into the Gaussian fitting procedure.}
    \label{fig:zpve}
\end{figure}

Fig. \ref{fig:zpve} shows the BE$_d$ for the two different surfaces of ASW, representing the core results from this article. The CO-\textit{np}ASW and CO-\textit{p}ASW BE$_d$s are comprised of 415 and 674 unique binding sites, respectively.  
A statistical analysis of the quartiles of the distributions, suggests the definition of 
5 intervals of BE, namely Very Low (VL), Low, Medium, High, Very High (VH) (see Sec. \ref{sec:methods_be} for details). 
Both BE$_d$ have a similar mean BE of around 1100 K which is lowered significantly by adding the ZPVE correction to 889 K for 
the \textit{np}ASW surface and 911 K for \textit{p}ASW. 
It is thus important to highlight 
that the existence of the nanopores does not notably affect the statistical BE profile of CO on ASW. One notable 
difference however is the longer high-BE tail in the \textit{p}ASW which is also reflected in a higher maximum BE value of
1845 K vs 1662 K in the case of \textit{np}ASW.
The significant standard deviations ($\sigma$: 277 K and 292 K for \textit{np}ASW and \textit{p}ASW, respectively) relative 
to the mean BE values reflect the wide range of available binding sites on the realistic, amorphous surfaces. 
This structural diversity enables the CO molecule to explore an extensive variety of adsorption configurations, 
effectively sampling the full spectrum of binding modes possible, which, put together, results in a broad distribution. The similarity of both BE$_d$s  is somewhat surprising as the \textit{np}ASW 
and \textit{p}ASW surfaces exhibit different features in terms of nanopores and number of 
dangling-OH bonds. Therefore, in the next section  we try to rationalize these results using SAPT analysis of the binding sites. 

\begin{table}[h]
\begin{center}
\caption{Percentages (\%) and average (avg) BE for each group defined in Sec. \ref{sec:methods_be}: 
}
\label{tab:avgBE}
\begin{tabular*}{0.49\textwidth}{@{\extracolsep{\fill}}lccccc}
\toprule

System & BE groups  &  \% & min & avg & max\\ 
       \bottomrule
\multirow{5}{*}{CO-\textit{np}ASW}  &    VH  
            &  5       & 1333 & 1454 &  1662            \\
  &     High & 14 & 1111 & 1185 & 1333   \\
  &     Medium & 62 & 667 & 883  & 1111\\
  &     Low & 14 & 444 & 577 & 667   \\
  &     VL & 5 & 50  & 331  & 444    \\
 \bottomrule
 \multirow{5}{*}{CO-\textit{p}ASW}  &    VH  
             & 4 & 1367 & 1489 & 1845 \\
  &     High & 16 & 1139 & 1239  & 1367 \\
  &     Medium & 58 & 683 & 922 & 1139\\
  &     Low & 17 & 456 & 586 & 683   \\
  &     VL & 5 & 107 & 350 &  456   \\
  \bottomrule
\end{tabular*}
\tablefoot{
Very Low (VL), Low, Medium, High, Very High (VH). BE values in K.
}
\end{center}
\end{table}

\subsubsection{Interaction energy decomposition distribution}
\begin{figure*}
    \centering
\includegraphics[width=\linewidth]{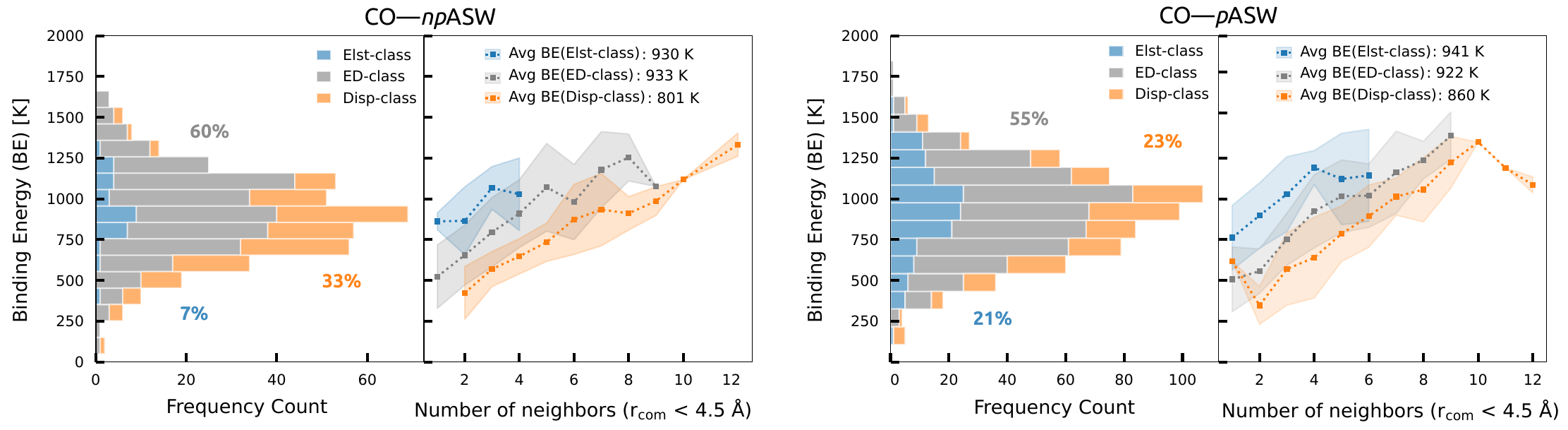} \\
    \caption{CO-\textit{np}ASW panel:   
Left: BE distribution for CO on \textit{np}ASW  ice, with color mapping highlighting the dominant interaction contributions 
    at various binding sites, ranging from  primarily electrostatic (Elst-Class, blue) to purely dispersion (Disp-Class, orange), while ED-Class (grey) stands for an intermediate group.  
    The coefficient that determines to which class the binding site belongs is 
    the ratio between the dispersion energy and the sum of the attractive interaction energies (electrostatic, induction and dispersion), 
    and is defined in Sec. \ref{sec:ie}.  
  Right: correlation between the BE and CO number of neighbors (i.e. number of water molecules in a radius of 4.5 \text{\AA}  from CO center of mass) for each class of binding sites.  Averages are represented as points connected by a dashed line and the standard deviation is shown as a shaded region of matching color. 
 CO-\textit{p}ASW panel: analogous for \textit{p}ASW ice.}
    \label{fig:be_sapt}
\end{figure*}

In order to explain the similarities of the BE$_d$ of CO adsorbed on \textit{np}ASW 
and \textit{p}ASW, we estimated the interaction energy 
contributions for 
the totality of the binding sites. Due to system size limitations, we 
conducted a SAPT analysis
considering the 28-water molecules closest to CO's center of mass in each binding site -  
the size of the extracted clusters has been chosen according to our test, which showed that the different contributions of the interaction energy are converged after 22-25 molecules; see also Appendix Fig. \ref{fig:sapt_incr} for an example of the convergence of two structures.
The analysis aims   
to statistically evaluate the  characteristics of the interaction of CO with the two types of ASW surfaces;  
from the interaction energy contributions we also evaluated $F_\mathrm{disp}$, Eq. \ref{eq:F_disp}), whose distribution is reported in  
Fig. \ref{fig:be_sapt}, left panels, for the two ASW models, 
mapped on top of 
the bins of the ZPVE-corrected BE$_d$s. 
For both the BE$_d$s, CO is  
bound by an interplay of electrostatic and dispersion interactions in the majority of the binding sites. The structures corresponding to this group (ED-class, gray bins) are present across all 
BE regimes. One noticeable difference between both surface
types is that in \textit{np}ASW there is a higher percentage of structures
that are bound primarily by dispersion (Disp-class, orange bins, 33\%) and less by electrostatic 
interactions (Elst-class, blue bins, 7 \%), while in \textit{p}-ASW both groups are present in a similar 
proportion (23\% vs 21\% for Elst- and Disp-class,  respectively).  This difference can be attributed to the  surface morphology (Sec. \ref{sec:results:surface}): 
the \textit{np}ASW surfaces contain fewer dangling-OH bonds, 
and more small surface gaps, where the CO
can bind effectively through dispersion interactions.
On the other hand, the \textit{p}ASW has more crests and valleys, which results in a larger 
proportion of dangling-OH that  facilitates
electrostatic coupling. 
 
\begin{figure}
    \centering
    \includegraphics[width=\linewidth]{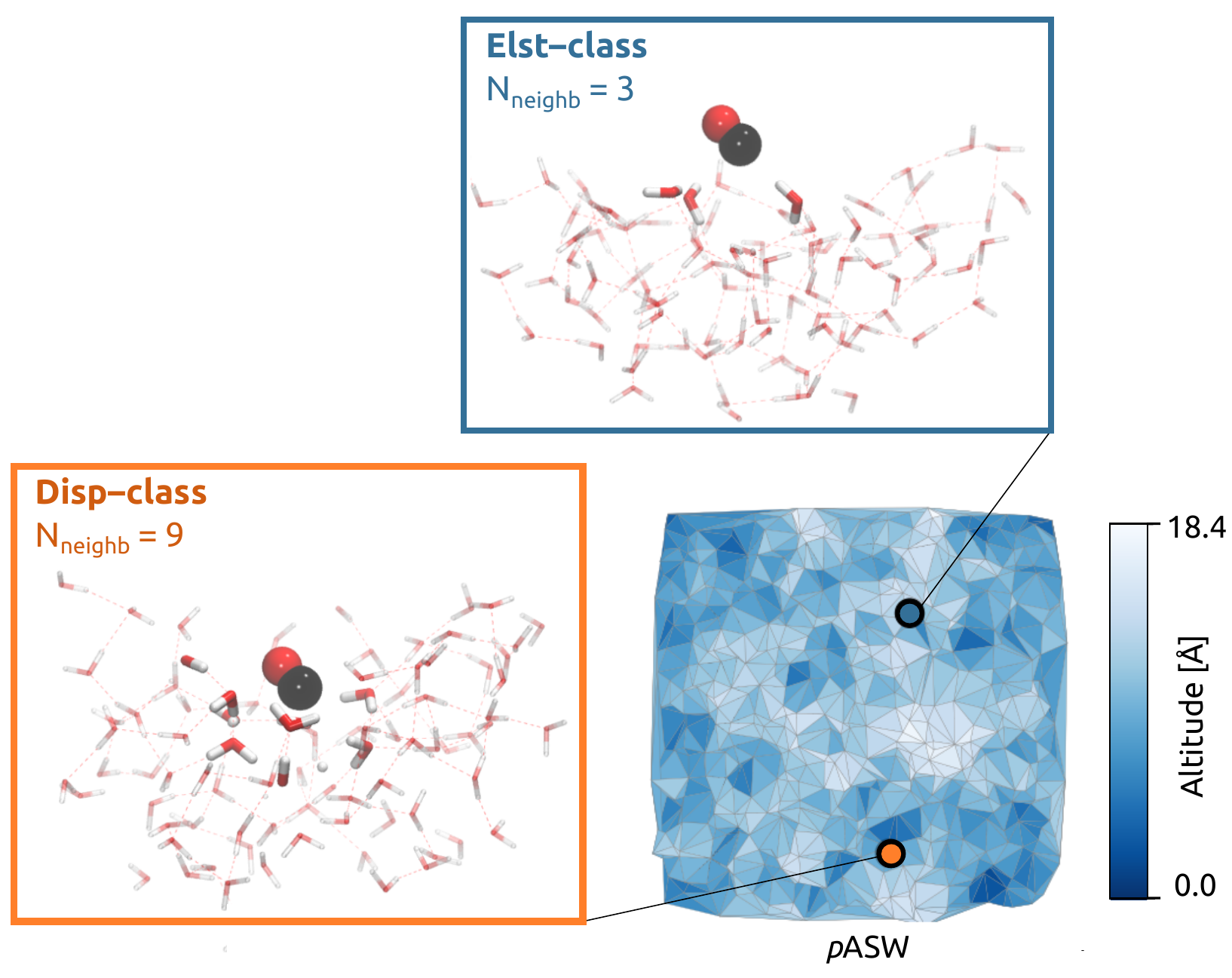} 
    \caption{Example of Very High (VH)-BE structures belonging to Elst-class (blue, BE = 1370 K) and Disp-class (orange, BE = 1492 K), and their location on the \textit{p}ASW surface, represented as altitude map. 
    The inset figures  display a portion of the binding sites comprising 50 water molecules, represented as sticks. The water molecules within 4.5 $\text{\AA}$ of CO center of mass (i.e. nearest neighbors, N$_{neighb}$) have been highlighted.}
\label{fig:example_classes}
\end{figure}

The Elst-class displays the highest average BE for the \textit{p}ASW (941 K), whereas in \textit{np}ASW both ED-class and Elst-class have practically the same value of 930 K. 
For both type of surfaces, the structures  bound by dispersion
(Disp-class) are mostly  located in the center and low BE part of 
the distribution, while
the high-BE tail (VH-BE group, violet bar in Fig. \ref{fig:zpve}) 
is predominantly populated 
by ED-class 
structures.  
In order to shed light on the characteristics of the VH-BE ED-class structures, we first analyzed two representative VH-BE structures bound mainly by electrostatics or dispersion. 
The selected examples are reported in the insets of Fig. \ref{fig:example_classes};  since 
they derive from the same ice model, the figure also exhibits their location (we remind the reader that each binding site is optimized individually, and the figure just overlap their ice altitude maps). 
The VH-BE Elst-class (BE = 1370 K), blue inset,  
is characterized by the   
presence of a water molecule acting as H-bond to the C-extremity of the adsorbate. Within this binding site, the two water molecules closest to the CO molecule do not engage in hydrogen bonding with one another, reflecting a configuration akin to that observed in \COtrc.
In fact, 
this binding site shares similar structural parameters with \COtrc, such as   
bond distances, as well as $F_\mathrm{disp}$ value. Interestingly, such binding motive on the ASW surface is enabled 
by the irregularities peculiar to the ASW framework, and therefore, the adsorption process does not require the 
large amount of deformation energy as in \COtrc, resulting in a 
very favorable binding site, with a non ZPVE-corrected BE 
similar to \COtrc interaction energy.  
In fact, as demonstrated in the example provided 
in Fig. \ref{fig:example_classes}, there are only 
three water molecules within a radius of 4.5 $\text{\AA}$ 
from the center of mass of the CO, thus resembling a water trimer.\\
In order to extend this observation
to other structures in Elst-class, we quantified 
the number of nearest neighboring water molecules for each 
binding site. This was determined by counting the water molecules 
within a radius of 4.5 $\text{\AA}$ from the center of mass of 
the CO molecule ($N_\mathrm{neigh}$). The outcomes of this 
analysis are presented in the right panels of Fig. \ref{fig:be_sapt}.
For Elst-class interactions on both surfaces, there are significantly 
fewer water molecules surrounding the binding sites, compared to the 
other interaction types, reaching a maximum of $N_\mathrm{neigh}$: four for \textit{np}-ASW
and six for \textit{p}-ASW. This is expected since, for electrostatic 
interactions to be dominant, a particular arrangement of water molecules 
 needs to be present,  
the possibility of having binding sites crowded with water molecules.\\
On the other hand, the binding environment for Disp-class molecules is quite different. An example 
is shown in the orange inset of Fig. \ref{fig:example_classes}. For this structure the CO molecule is
``floating'' on the bottom of a concave ice portion, without establishing strong interactions to any single 
water molecules, and thus dispersion interaction to the ASW surface dominates. Such arrangement is, naturally,
not accounted for in the small water clusters (Fig. \ref{fig:be_model}), and provides a exceptionally high dispersion energy ($E_\mathrm{disp}$) value (2600 K, see Appendix Table \ref{tab:sapt_model} for comparison), benefiting from the proximity 
of several water molecules ($N_\mathrm{neigh}$: 9). 
A more general inspection of this interaction type compared to the Elst-class can be seen in 
the right panels of Fig. \ref{fig:be_sapt}. While Disp-class binding sites generally exhibit lower BEs at low coordination (i.e., when few water molecules are present within a 4.5 \text{\AA} radius), their BE increases significantly as surrounding water molecules accumulate. This provides direct evidence supporting the long-held intuition that stronger binding occurs in confined cavities. Nevertheless, the incremental SAPT analysis presented in Appendix \ref{sec:ap_sapt} Figure \ref{fig:sapt_incr} 
reveals that the contribution of E$_\mathrm{disp}$ reaches near convergence upon the inclusion of 9 to 11 water molecules. 
Additional water molecules situated at greater distances contribute only marginally to the overall dispersion energy.
Finally, the ED-class structures present a combination of the characteristics of 
both Elst- and Disp-classes, such as the favorable interaction given by the establishment of H-bonds with the ice, and the additive nature of the dispersion contribution to the BE with the number of surrounding water molecules. 
In fact, a typical VH-BE ED structures is highly embedded in the ASW network- as the adsorbate is interacting 
with the substrate forming H-bonds- and it is located in a valley region, therefore benefiting from 
the closeness of several water molecules. A representative example is displayed in Appendix Fig. \ref{fig:ed_class}.\\            
In summary, despite the average BE values and the shape of the BE$_d$ being comparable
for the CO-\textit{np}ASW and CO-\textit{p}ASW systems, a closer inspection of the 
nature of the interactions between CO and the surfaces reveals noticeable differences 
arising from their distinct morphological characteristics. Specifically, the higher abundance 
of dangling-OH bonds on \textit{p}-ASW promotes a comparatively stronger electrostatic binding
regime (Elst-class) via hydrogen bonding with the substrate. 
Furthermore, the high-BE tail of the BE$_d$ in both ASW types is predominantly 
composed of ED-class structures, which benefit significantly from the presence 
of dangling-OH bonds in conjunction with additive dispersion interactions due to 
the close proximity of multiple water molecules to CO. The in-depth analysis presented in this section aims to set a stepping stone for future studies of adsorbates on \textit{p}ASW. Although we find similar BE$_d$s,
it is likely that CO represents a special case among the broad spectrum of possible adsorbates, due to the delicate interplay between electrostatic and dispersion interactions. We anticipate that more polar adsorbates will exhibit stronger binding to \textit{p}ASW, attributed to the higher density of dangling-OH bonds in this model. If confirmed, this behavior could have significant implications for the segregation of polar and apolar molecules on ice surfaces. We intend to explore this line of research further in future work.

\subsection{Simulated TPD curves} \label{sec:discussion:literature}

Our detailed BE map allows us to generate realistic TPD curves akin to 
those observed in  multicoverage TPD experiments. To this end, it is essential
to take into account that CO molecules are capable 
of diffusing as the temperature increases prior to the onset of desorption.
The  extent of the diffusion depends on the initial CO coverage of the 
substrate: at full coverage, diffusion is restricted as all binding sites
are occupied. However, as initial coverage decreases, molecules gain the ability 
to diffuse into empty binding sites, especially those with higher BE.
Consequently, desorption predominantly occurs from these higher-BE sites. Having
a complete BE distribution, we can simulate various scenarios 
corresponding to different degrees of coverage. Specifically, by applying 
progressively left-truncated distributions—excluding low BE sites—we can
approximate desorption behavior under different conditions, ultimately 
analyzing scenarios in which only high-BE sites remain available. 
In total, we computed TPD traces considering four different truncations 
of the BE$_d$,  and two pre-exponential factors,
following the procedure explained in Appendix \ref{sec:ap_TPD}. We used a 
heating rate of 0.3 Ks$^{-1}$ which lies in the range commonly used 
in multicoverage TPD experiments (0.2-0.5 K s$^{-1}$). 
The initial coverage ($\theta_i$) corresponds to the fraction of the remaining
binding sites after truncation of the BE$_d$. 
Fig. \ref{fig:TPD} shows the results for CO-\textit{np}ASW (top panel) and CO-\textit{p}ASW (bottom panel). 
The solid lines and  segmented  lines correspond to TPD traces derived with a  
pre-exponential factor of $\nu = 10^{12}$ s$^{-1}$ 
and $\nu = 9.14\times 10^{14}$ s$^{-1}$, respectively. 
Simulated scenarios  with lower coverages 
(i.e., corresponding to left-truncated distributions with higher BEs) 
yield TPD traces which are shifted to higher temperatures  and with lower desorption flux, 
consistent with experimental TPD traces. The TPD peaks for the high coverage case lie  
around 30 K, while for the low coverage cases they are located around 50 K. 
The higher pre-exponential factor (segmented line)  shifts the TPD traces by
8 to 9 K to lower temperatures,  consistent with a faster desorption kinetics associated to an increased desorption attempt frequency.
The TPD traces from all the truncations exhibit a common tail toward higher temperatures, 
which is expected, as they all stem from the same high-BE bins desorption flux. This is noteworthy, as it is a feature also observed in multicoverage TPD experiments. Finally, the results for CO-\textit{np}ASW and  
CO-\textit{p}ASW  closely resemble each other, mirroring the results of their similar BE$_d$. 

\section{Discussion} \label{sec:discussion}
\subsection{Comparison with experimental TPD results} \label{sec:discussion:literature}

\begin{figure}
    \centering
    \includegraphics[width=\linewidth]{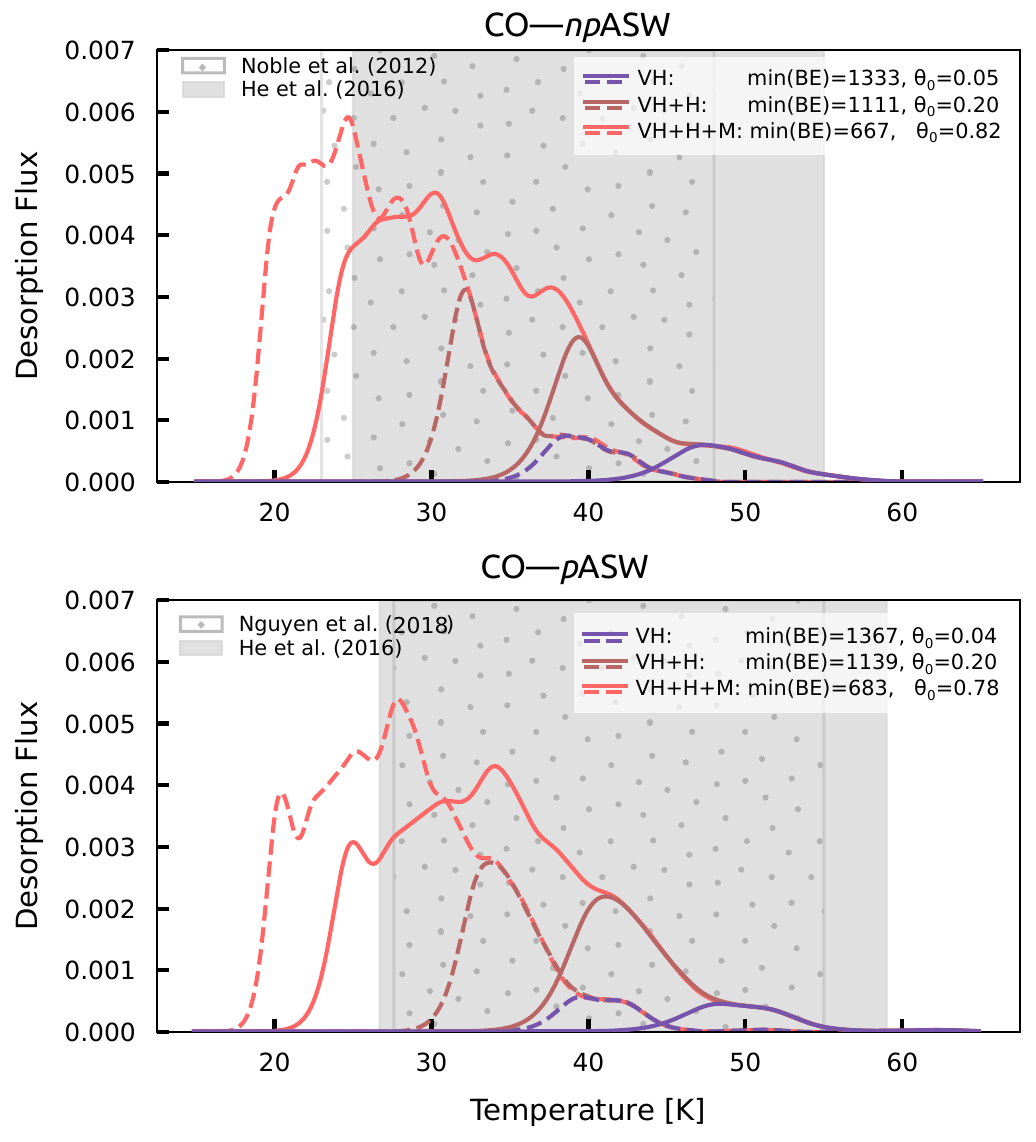} \\
    \caption{Simulated TPD traces using data from our BE distributions compared to different experimental results of multicoverage TPD desorption from 
    \cite{noble_thermal_2012}, \cite{nguyen_segregation_2018} and \cite{he_binding_2016}. Heating ramp rates ($\beta$) is 0.3 K s$^{-1}$. 
    (See Appendix \ref{sec:ap_TPD} for further details).
    The different colored TPD traces are simulated considering different BE cutoff in the distribution, meaning that BEs lower than min(BE) where not 
    considered in the simulation (see Table \ref{tab:TPD peaks}). The initial coverage of each TPD trace corresponds to the fraction of molecules of the full distribution that remains after the cutoff is applied. The shaded area represents the range
    of the experimental TPD. TPD traces with two different pre-exponential factors are shown: $\nu = 10^{12}$ s$^{-1}$  (solid line) and $\nu=9.14\times10^{14}$ s$^{-1}$ (segmented lines). The 
    latter is obtained from transition state theory as derived 
    in \cite{minissale_thermal_2022}.
    }
    \label{fig:TPD}
\end{figure}

In order to asses the relation between our BE$_d$s and the experimental 
BEs that were derived from TPD experiments, we directly compared 
our simulated TPD traces to three different experiments  
\citep{noble_thermal_2012, nguyen_segregation_2018, he_binding_2016}, that 
obtained multicoverage TPD traces for CO bound to different ASW types. The  temperature regions, 
corresponding to the desorption fluxes observed from these experiments, are depicted as 
shaded areas within Fig. \ref{fig:TPD}. The  simulated traces fall mostly within the   
experimental ranges, across all experiments. In both the experimental
and theoretical TPDs, the signals corresponding to \textit{p}ASW are shifted to higher temperatures compared to \textit{np}ASW, which is in agreement with the longer 
high-BE tails observed in the CO-\textit{p}ASW distribution. 
However,
for the high coverage cases, the simulated TPD traces are
slightly shifted to lower temperatures compared to the experimental results.
A reason for the disagreement might be the lateral CO-CO interactions that 
are not present in our simulated TPD traces and could affect desorption 
profile at that regime. 

The pre-exponential factor has a noticeable effect 
on the position of the signals in the simulated TPD-spectra. 
The pre-exponential factor derived from Transition State Theory (TST) ($\nu = 9.4\times10^{14}$ s$^{-1}$ \citep{minissale_thermal_2022}) 
shifts the TPD traces (segmented lines in Fig. \ref{fig:TPD}) to lower temperatures, 
which leads to a poorer agreement with the experimental results.
When discussing comparison between simulated TPD curves and experimental ones,
it is important to mention that \cite{bariosco_binding_2024} recently
generated simulated TPD traces derived from  \textit{ab initio} \ce{H2S} BE$_d$ on a large cluster ASW model.  In their work, they also detected 
a shift to lower temperatures for the simulated TPD with respect to the experimental one. 
However, the shift is much larger than the one observed in this work, which is puzzling 
since the diffusion of \ce{H2S} should be less than in the case of CO, as the molecule can 
interact through hydrogen bonding to the surface, and has a much higher interaction energy with the ASW surface. The disagreement with experimental results
in the case of \ce{H2S} simulated TPD traces, most likely stems from an overestimation of the ZPVE correction of the BE values, which 
was computed for every binding sites using embedding procedures, potentially contributing to the discrepancy.

\subsection{Comparison with theoretical binding energy values}

Over the years, many attempts have been made to compute the CO - water interaction 
using force fields or quantum chemistry methods.  \cite{karssemeijer_dynamics_2013}
studied the interaction of one to six CO molecules on ASW using classical
force fields parametrized using high-level \textit{ab initio} data of 
the \ce{H2O}-\ce{H2O} and \ce{H2O}-\ce{CO} dimers. Similarly to  the present work, 
they obtained a BE$_d$, albeit shifted to higher energies
(650 K - 2900 K). The difference might be ascribed to the inability of force-fields to 
fully capture the many-body polarization and subtle orientation-dependent 
interactions in complex, disordered environments like ASW. Considering that the interactions in force-fields are
 parametrized on ideal model systems, the resulting BE values are naturally higher 
than the ones reported in the present work, obtained using a MLP that is trained on a
myriad of realistic ASW like clusters configurations. The only other BE$_d$  
for a CO-ASW was presented in \cite{bovolenta_binding_2022}, employing
a set of amorphized water clusters containing 22 molecules each, serving as 
surface model. Using a $\omega$-PBE/def-TZVP//M05/def-TZVP level of theory,  
together with BSSE correction, yielded a non-ZPVE corrected average BE of 1035 K, 
which is slightly lower than the one reported in this study. However, the standard deviation
of the set-of-clusters model BE distribution was significantly lower, with a value of 176 K vs 277-292 K 
reported in this work. Considering that such medium-sized clusters have a higher proportion 
of dangling-OH bonds compared to both \textit{np}ASW and \textit{p}ASW models, the narrower distribution can be
rationalized by examining the Elst-class binding sites defined in this study, as Elst-class sites 
are likely predominant in the cluster model. Within this class, the spread of BEs is considerably lower
(See Fig. \ref{fig:be_sapt}, blue bins), while the average BE remains similar to the total average BE$_d$ and 
across all three classes.
Therefore, the 
inclusion of dispersion-dominated structures appears to affect the spread of the BE$_d$, and the average to a lesser extent. Hence, a geometrical   
 similarity 
 between the set-of-cluster binding sites and the Elst-class sites would explain an analog similarity in the the average BE. 
A different approach was
used in the  study of \cite{ferrero_binding_2020} in which they obtained a range of BEs 
by placing the CO on different sites of an amorphous water slab consisting of 
60 water molecules. They used a periodic systems at the  B3LYP-D3/A-TZV level 
of theory for the BE  computation on a HF-3c optimized binding sites. The non-ZPVE-corrected BE range of 1300 K to 2200 K, falls within the high BE half of the BE$_d$s presented in this study. This tendency is in agreement with our benchmark, which showed that the 
B3LYP-D3BJ method tends to overbind the small reference systems (see Appendix \ref{sec:ap_DFTbench}) — potentially contributing to the higher BEs.
Additionally, the HF-3c level of theory for binding site optimization may not be sufficiently reliable 
for accurately describing CO–W interactions \citep{bovolenta_binding_2022}. Concurrently with our
study, \cite{Groyne_robust_2025} obtained a BE$_d$ on 
ASW using a multi-level approach. They first built an ASW model
with a classical force field, from which they extracted and 
sampled a hemisphere using ONIOM2 embedding. In their scheme,
the high-level zone (the binding site and molecules within 8~\text{\AA})
was treated at the B3LYP-D3BJ/6-311G(d,p) level of theory, 
while the low-level region was handled by the GFN2-xTB method. 
Their ZPVE and BSSE corrected  BE is higher than 
the one presented here, amounting to a average value of 
1400 K and a standard deviation of 340 K. The significantly 
higher values in their work are puzzling as the ASW 
surface prepared in their work is similar to the one presented here. A source of discrepancy could be the level of theory: as 
mentioned above, B3LYP-D3BJ 
tends to overestimate the BE of CO + ASW. Another possible source of discrepancy is the polarization from the outer, GFN2-treated, region in the ONIOM2 calculation. This term contributed up to 25\% of the total BE, a surprisingly large influence, given the considerable size of the high-level zone. In contrast, our own SAPT0-D3BJ incremental analysis demonstrates that more distant molecules have a negligible effect on the interaction energy. Further studies comparing the two  approaches will be needed to reconcile these differing results.

\begin{table}[h]
\begin{center}
\caption{BEs and TPD peaks for different binding energy regimes.} 
\label{tab:TPD peaks}
\begin{tabular*}{0.48\textwidth}{@{\extracolsep{\fill}}llcc}
\toprule
\small
System & BE regime & min(BE),  Avg(BE) & TPD peak  \\ 
       \bottomrule
\multirow{4}{*}{CO-\textit{np}ASW}
  &    VH  & 1333,  1454  & 47 (38) \\
  &     VH + H & 1111,   1258 & 39 (33)  \\
  &     VH + H + M & 667,  973 &  31 (25)\\
 \bottomrule
\multirow{4}{*}{CO-\textit{p}ASW}
  &    VH  & 1367, 1489   & 48 (39) \\
  &     VH + H & 1139, 1288 & 41 (33)\\
  &   VH + H + M & 683, 1017 & 34 (28)\\
  \bottomrule
\end{tabular*}
\tablefoot{Minimum (min), average (Avg) BE values and TPD peaks corresponding to  
the three truncated
BE regimes: Very high (VH), High (H) + VH and Medium (M) + H + VH. The heating rate 
to obtain the simulated TPD trace is 0.3 K s$^{-1}$ and the pre-exponential
factor $\nu$ = $10^{12}$ ($9.14\times10^{14}$) s$^{-1}$. }
\end{center}
\end{table}

\subsection{Protoplanetary disk's CO snowlines from our derived distributions}

One of the key quantities used to characterize the distribution of molecules in protoplanetary disks (PPDs) is the so-called snowline, typically defined for \ce{H2O}. This marks the region in the disk where water vapor condenses into ice, setting the boundary between the formation zones of rocky and gas-giant planets \citep{notsu_water_2020}. In this context, predicting the corresponding snowline for CO becomes essential for understanding the distribution of organic molecules in PPDs, as well as the organic inventory available to forming planets.

\begin{figure*}
    \centering
    \includegraphics[width=0.95\linewidth]{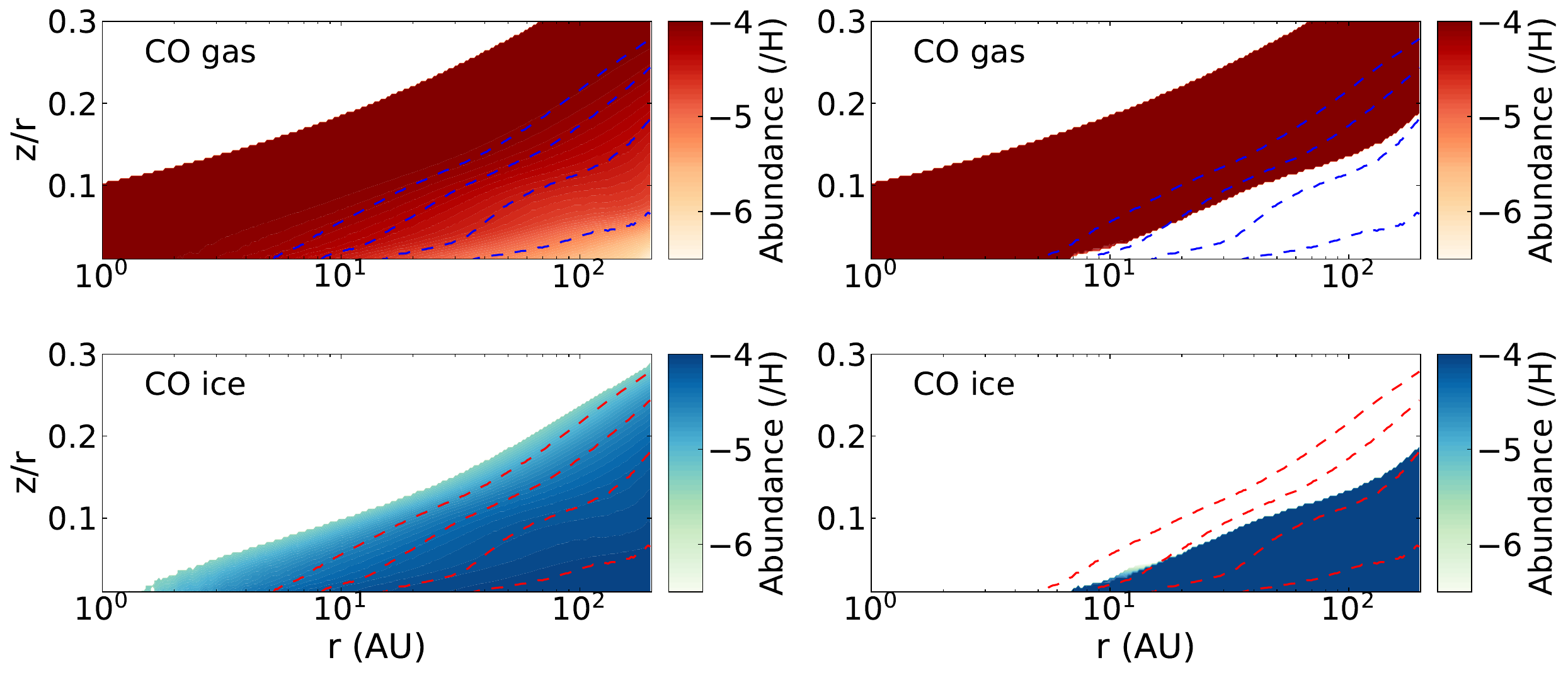} \\
    \caption{2D spatial distributions of CO gas (top panels) and ice (bottom panels) abundances on the disk using a multibinding description (left panels) and single binding description (right panels). The vertical axes represent height normalized by the radius. The dashed lines depict the positions where the dust temperature is equal to 30 K, 25 K, 15 K, and 10 K.}
\label{fig:model2}
\end{figure*}

\begin{figure*}
    \centering
\includegraphics[width=0.93\linewidth]{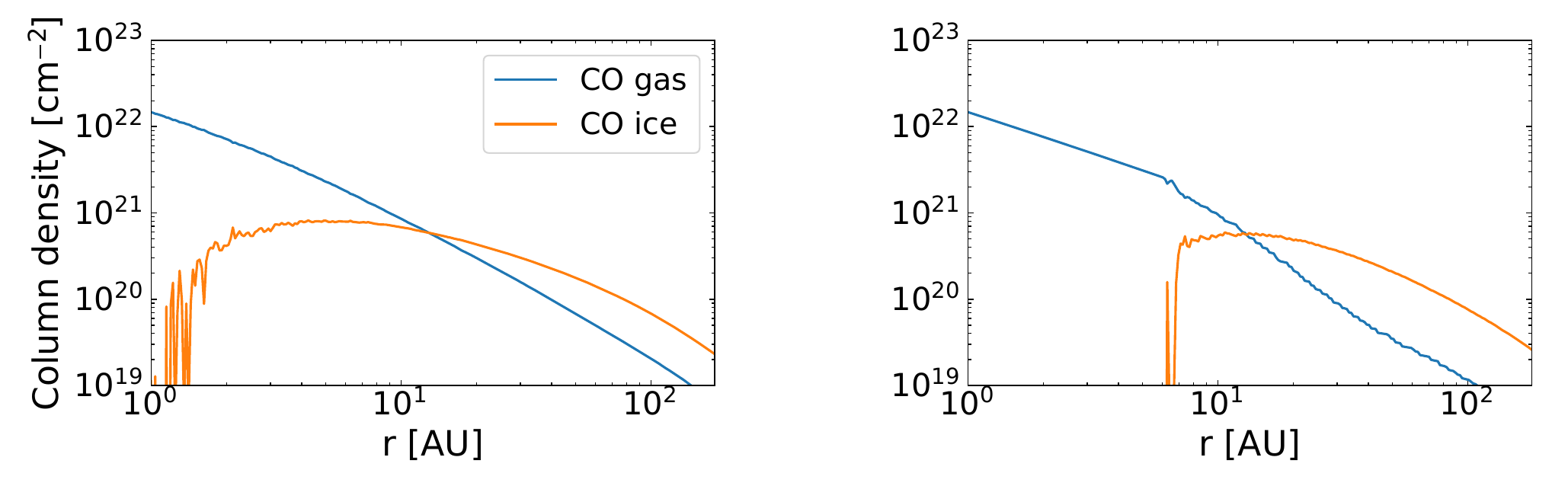} \\
    \caption{Vertically integrated CO gas and ice column densities as functions of the radius. The left panel shows the model with the multibinding description, while the right panel shows the single binding description.}
\label{fig:model1}
\end{figure*}

To contextualize our results under this prism, we simulated the partitioning of CO between the gas-phase and the solid-phase at each position of a 2D protoplanetary disk model.
We use the density and temperature profiles for gas and dust and the stellar UV radiation field appropriate for the TW Hya disk \citep{cleeves_constraining_2015,furuya_detection_2022}.
We assume the total (gas + solid) CO abundance with respect to H nuclei of 10$^{-4}$ in the whole disk.
In the disk atmosphere, however, the strong UV radiation from the central star dissociates CO.
Then we assume no CO molecules exist in regions with $A_V < 1$ mag \citep{aikawa_nomura_2006}.

The partitioning of CO between the gas phase and the solid phase is calculated in two ways.
In a first approach, we use our BE$_d$ of CO on \textit{np}ASW and calculate the partitioning in a similar way to that in \citet{tinacci_theoretical_2023};
at each disk position, we calculate the fraction of adsorption sites where $k_{\rm thdes} > k_{\rm ads}$ ($f_{\rm gas}$), and the gas-phase and solid-phase CO abundance at each disk position are given by $10^{-4}f_{\rm gas}$ and $10^{-4}(1-f_{\rm gas})$, respectively.
Here $k_{\rm thdes}$ and $k_{\rm ads}$ are the rate constants for thermal desorption and adsorption of CO, respectively.
In a second approach, we compare $k_{\rm thdes}$ evaluated with the mean binding energy of CO on \textit{p}-ASW with $k_{\rm ads}$ at each position of the disk. When the former is larger (smaller) than the latter, we assume all CO molecules at the position exist as vapor (ice).
In both cases, the pre-exponential factor $\nu$ for $k_{\rm thdes}$ is set to 10$^{12}$ s$^{-1}$. We also tried using a $\nu$=9.14$\times$10$^{14}$ s$^{-1}$, not obtaining significant changes.

Fig. \ref{fig:model2} shows the 2D (radial + vertical) distribution of the gas-phase and the solid-phase CO using the multibinding description (right) and the single binding description (left).
In the single binding description, the partitioning of CO between gas and solid phases dramatically changes around the temperature of 20-25 K, above (below) which CO is predominantly present in the gas (solid) phase, i.e., the CO snowline (or snow surface) can be defined as the temperature of 20-25 K.
On the other hand, in the multibinding description, non-negligible fraction of CO exists in the gas phase even at 15 K, while non-negligible fraction of CO exists in the solid phase even at 30 K.
Therefore, the CO gas and ice coexist in larger regions of the disk in the multibinding description than in the single binding description.
This can be seen in the vertically integrated CO gas and ice column densities as shown in Fig. \ref{fig:model1}. 

The change in snowlines anticipated by our results carries important implications for the chemistry of PPDs. Because CO is considered the most important precursor of complex organic molecules through reactions on ice \citep{Watanabe2002}, changes in this feedstock affect the overall distribution of organics in PPDs. A multibinding approach suggests a broader spatial extension of CO in the disk midplane and beyond, along with a corresponding enhancement in its gas-phase distribution, driven by the presence of both high- and low-binding-energy sites. This, in contrast to the predictions of a single-binding approach, points to a richer CO chemistry at low stellar radii. Finally, our results reproduce the trend predicted in the results of \citet{grassi_novel_2020} for CO partitioning using BE distributions, providing a quantitative estimation of their effect.

\subsection{Implications for other astrophysical regions}\label{sec:discussion:astro}

Considering a multibinding approach to the BE has direct implications for PPDs, as discussed above. In the context of cold molecular clouds, however, the implications differ. At ultracold temperatures ($\sim$10 K), adopting a multibinding approach has a negligible effect on the desorption rates (Fig. \ref{fig:TPD}); nonetheless, it can significantly influence CO diffusion, and consequently, the formation of CO-related species. This has already been demonstrated in previous works by some of us \citep{molpeceres_enhanced_2024, furuya_framework_2024}, particularly for reactions involving CO other than hydrogenations, most notably in the case of \ce{CO2} formation. In warmer regions like hot cores, the multibinding approach is expected to have little impact, as CO is not significantly depleted onto dust grains under these conditions.

\section{Conclusions} \label{sec:discussion:conclusions}
In this paper, we presented the binding energy distribution of CO bound to a porous and non-porous 
amorphous solid water surface. We used a machine-learning potential that was trained using forces
and energies of CO interacting with differently sized clusters (\ce{W_22} - \ce{W_60}) computed at the MPWB1K-D3BJ-gCP/def-TZVP. The main results from this study can be summarized as follows: 

\begin{enumerate}
    \item 
    Quantifying the interaction of CO on small water clusters at high level of theory (CCSD(T)/CBS) allowed to construct a DFT benchmark to obtain a suitable model chemistry for the training set of the MLP. In small clusters, BEs span approximately between 
300 and 1500 K, generally decreasing when additional water molecules form more rigid hydrogen-bonded 
networks. Symmetry-Adapted Perturbation Theory (SAPT) analyses highlight the dominance of 
electrostatic and induction interactions, while dispersion contributes less significantly in 
these small systems.

\item An analysis of the compact non-porous (\textit{np}ASW) and porous (\textit{p}ASW) surfaces revealed  significant morphological 
differences between them. The porous surfaces exhibit greater roughness, deeper valleys, and 
higher densities of dangling-OH bonds, compared to the smoother non-porous surfaces. Quantitative
metrics such as areal average roughness and mean roughness depth confirm these structural 
differences. 

\item The binding energy distributions obtained for these systems exhibit Gaussian-like shapes 
with mean zero-point vibrational energy-corrected values near 900 K, showing little overall
difference between porous and non-porous surfaces. Interaction energy decomposition, however,
reveals important differences: \textit{p}ASW favors electrostatic interactions due to the 
higher density of dangling-OH groups, while \textit{np}ASW  exhibits a greater proportion 
of dispersion-dominated binding sites. High-BE sites typically involve a balanced contribution
from both electrostatic and dispersion interactions, highlighting the intricate interplay 
between local morphology and molecular-level interactions. These findings emphasize the 
complexity and diversity of CO adsorption on ASW, relevant for accurately 
modeling astrochemical processes. The limited variability in the binding energy distributions of CO on \textit{np}ASW and \textit{p}ASW, resulting from a balance between electrostatic and dispersion interactions, suggests that the observed independence from substrate morphology is a unique characteristic of CO and perhaps other apolar molecules, like \ce{CO2} or \ce{CH4}. This behavior is unlikely to extend to polar adsorbates, such as \ce{NH3} or \ce{CH3OH}, which are expected to exhibit binding energy distributions shifted toward stronger binding on \textit{p}ASW, due to the uneven distribution of dangling-OH bonds.

\item We proposed a new approach to compare full binding energy distributions to experimental 
TPD studies based on truncation of the  distribution to mimic a low-coverage
scenario and diffusion to high-binding energy sites before desorption. Using this approach, we 
effectively reproduce the main features observed experimentally in multicoverage TPD curves of CO on ASW surfaces. We find that our simulated TPD traces mostly fall within the 
experimental peaks range (25-55 K).

\item Compared to single-value models, our multibinding-energy approach results in a
broader and more diffuse CO snowline in protoplanetary disks, significantly extending the region where CO 
ice and gas coexist. This expanded coexistence zone, spanning approximately 15–30 K,
indicates more complex partitioning between gas and ice phases, which carries critical
implications for chemical evolution and organics delivery in planet formation scenarios within 
protoplanetary disks.

\end{enumerate}

We aim to expand this novel methodology for the computation of binding energies,
taking into account realistic statistical characteristics, inherent to ASW surfaces,
to other important interstellar molecules.

\section*{Data availability}
Molecular structures and corresponding energies and forces used for the training of the machine learning potential can be accessed online\footnote{\url{https://zenodo.org/records/16904328}}. Further information will be provided upon request by contacting the corresponding authors.

\begin{acknowledgements}
GMB gratefully acknowledges support from Proyecto UCO 1866 Beneficios Movilidad 2021–2022.
GM acknowledges the support of the grant RYC2022-035442-I
funded by MICIU/AEI/10.13039/501100011033 and ESF+. GM
also received support from project 20245AT016 (Proyectos Intramurales CSIC). We acknowledge the computational resources provided
by bwHPC and the German Research Foundation (DFG) through
grant no INST 40/575-1 FUGG (JUSTUS 2 cluster), the DRAGO
computer cluster managed by SGAI-CSIC, and the Galician Super-
computing Center (CESGA). The supercomputer FinisTerrae III and
its permanent data storage system have been funded by the Spanish
Ministry of Science and Innovation, the Galician Government, and
the European Regional Development Fund (ERDF).
SVG thanks VRID research grant 2022000507INV for financing this project. This work was funded by Deutsche Forschungsgemeinschaft
(DFG, German Research Foundation) under Germany’s
Excellence Strategy - EXC 2075 - 390740016. We acknowledge the support of the Stuttgart Center for Simulation Science
(SimTech).
\end{acknowledgements}

\bibliographystyle{aa}
\bibliography{aa55836-25.bib}

\begin{thebibliography}{86}
\expandafter\ifx\csname natexlab\endcsname\relax\def\natexlab#1{#1}\fi

\bibitem[{{Aikawa} \& {Nomura}(2006)}]{aikawa_nomura_2006}
{Aikawa}, Y. \& {Nomura}, H. 2006, \apj, 642, 1152

\bibitem[{Al-Halabi {et~al.}(2004)Al-Halabi, Fraser, Kroes, \&
  Van~Dishoeck}]{al-halabi_adsorption_2004}
Al-Halabi, A., Fraser, H.~J., Kroes, G.~J., \& Van~Dishoeck, E.~F. 2004, A\&A,
  422, 777

\bibitem[{Amiaud {et~al.}(2006)Amiaud, Fillion, Baouche, Dulieu, Momeni, \&
  Lemaire}]{amiaud_interaction_2006}
Amiaud, L., Fillion, J.~H., Baouche, S., {et~al.} 2006, J. Chem. Phys., 124,
  094702

\bibitem[{Bannwarth {et~al.}(2019)Bannwarth, Ehlert, \&
  Grimme}]{bannwarth_gfn2-xtbaccurate_2019}
Bannwarth, C., Ehlert, S., \& Grimme, S. 2019, J. Chem. Theory Comput., 15,
  1652

\bibitem[{Bariosco {et~al.}(2024)Bariosco, Pantaleone, Ceccarelli, Rimola,
  Balucani, Corno, \& Ugliengo}]{bariosco_binding_2024}
Bariosco, V., Pantaleone, S., Ceccarelli, C., {et~al.} 2024, MNRAS, 531, 1371

\bibitem[{Bossa {et~al.}(2014)Bossa, Isokoski, Paardekooper, Bonnin, Van
  Der~Linden, Triemstra, Cazaux, Tielens, \& Linnartz}]{bossa_porosity_2014}
Bossa, J.-B., Isokoski, K., Paardekooper, D.~M., {et~al.} 2014, A\&A, 561, A136

\bibitem[{Bovolenta {et~al.}(2020)Bovolenta, Bovino, Vöhringer-Martinez, Saez,
  Grassi, \& Vogt-Geisse}]{bovolenta_high_2020}
Bovolenta, G., Bovino, S., Vöhringer-Martinez, E., {et~al.} 2020, Mol.
  Astrophys., 100095

\bibitem[{Bovolenta {et~al.}(2024)Bovolenta, Silva-Vera, Bovino, Molpeceres,
  Kästner, \& Vogt-Geisse}]{bovolenta_-depth_2024}
Bovolenta, G.~M., Silva-Vera, G., Bovino, S., {et~al.} 2024, PCCP

\bibitem[{Bovolenta \& Vogt-Geisse(2025)}]{bovolenta_methyl_2025}
Bovolenta, G.~M. \& Vogt-Geisse, S. 2025, J Mol Model, 31, 104

\bibitem[{Bovolenta {et~al.}(2022)Bovolenta, Vogt-Geisse, Bovino, \&
  Grassi}]{bovolenta_binding_2022}
Bovolenta, G.~M., Vogt-Geisse, S., Bovino, S., \& Grassi, T. 2022, ApJS, 262,
  17

\bibitem[{Boys \& Bernardi(1970)}]{boys_calculation_1970}
Boys, S.~F. \& Bernardi, F. 1970, MolPh, 19, 553

\bibitem[{Bozkaya \& Sherrill(2017)}]{bozkaya_analytic_2017}
Bozkaya, U. \& Sherrill, C.~D. 2017, J. Chem. Phys., 147, 044104

\bibitem[{Chaabouni {et~al.}(2018)Chaabouni, Diana, Nguyen, \&
  Dulieu}]{chaabouni_thermal_2018}
Chaabouni, H., Diana, S., Nguyen, T., \& Dulieu, F. 2018, A\&A, 612, A47

\bibitem[{Cleeves {et~al.}(2015)Cleeves, Bergin, Qi, Adams, \&
  Öberg}]{cleeves_constraining_2015}
Cleeves, L.~I., Bergin, E.~A., Qi, C., Adams, F.~C., \& Öberg, K.~I. 2015,
  ApJ, 799, 204

\bibitem[{Clements {et~al.}(2018)Clements, Berk, Cooke, \&
  Garrod}]{clements_kinetic_2018}
Clements, A.~R., Berk, B., Cooke, I.~R., \& Garrod, R.~T. 2018, PCCP, 20, 5553

\bibitem[{Collings {et~al.}(2004)Collings, Anderson, Chen, Dever, Viti,
  Williams, \& McCoustra}]{collings_laboratory_2004}
Collings, M.~P., Anderson, M.~A., Chen, R., {et~al.} 2004, MNRAS, 354, 1133

\bibitem[{Cuppen \& Herbst(2007)}]{cuppen_simulation_2007}
Cuppen, H.~M. \& Herbst, E. 2007, ApJ, 668, 294

\bibitem[{Cuppen {et~al.}(2017)Cuppen, Walsh, Lamberts, Semenov, Garrod,
  Penteado, \& Ioppolo}]{cuppen_grain_2017}
Cuppen, H.~M., Walsh, C., Lamberts, T., {et~al.} 2017, Space Sci Rev, 212, 1

\bibitem[{Das {et~al.}(2018)Das, Sil, Gorai, Chakrabarti, \&
  Loison}]{das_approach_2018}
Das, A., Sil, M., Gorai, P., Chakrabarti, S.~K., \& Loison, J.-C. 2018, ApJS,
  237, 9

\bibitem[{Dohnálek {et~al.}(2003)Dohnálek, Kimmel, Ayotte, Smith, \&
  Kay}]{dohnalek_deposition_2003}
Dohnálek, Z., Kimmel, G.~A., Ayotte, P., Smith, R.~S., \& Kay, B.~D. 2003, J.
  Chem. Phys., 118, 364

\bibitem[{Duflot {et~al.}(2021)Duflot, Toubin, \&
  Monnerville}]{duflot_theoretical_2021}
Duflot, D., Toubin, C., \& Monnerville, M. 2021, Front. Astron. Space Sci., 8

\bibitem[{Dunning~T.H. {et~al.}(2001)Dunning~T.H., Peterson, \&
  Wilson}]{DunningT.H.2001}
Dunning~T.H., J., Peterson, K.~A., \& Wilson, A.~K. 2001, J. Chem. Phys., 114,
  9244

\bibitem[{Fedoseev {et~al.}(2015)Fedoseev, Cuppen, Ioppolo, Lamberts, \&
  Linnartz}]{fedoseev_experimental_2015}
Fedoseev, G., Cuppen, H.~M., Ioppolo, S., Lamberts, T., \& Linnartz, H. 2015,
  MNRAS, 448, 1288

\bibitem[{Fedoseev {et~al.}(2017)Fedoseev, Scirè, Baratta, \&
  Palumbo}]{fedoseev_cosmic_2017}
Fedoseev, G., Scirè, C., Baratta, G.~A., \& Palumbo, M.~E. 2017, MNRAS, 475,
  1819

\bibitem[{Ferrero {et~al.}(2020)Ferrero, Zamirri, Ceccarelli, Witzel, Rimola,
  \& Ugliengo}]{ferrero_binding_2020}
Ferrero, S., Zamirri, L., Ceccarelli, C., {et~al.} 2020, ApJ, 904, 11

\bibitem[{Fuchs {et~al.}(2009)Fuchs, Cuppen, Ioppolo, Romanzin, Bisschop,
  Andersson, van Dishoeck, \& Linnartz}]{fuchs_hydrogenation_2009}
Fuchs, G.~W., Cuppen, H.~M., Ioppolo, S., {et~al.} 2009, A\&A, 505, 629

\bibitem[{Furuya(2024)}]{furuya_framework_2024}
Furuya, K. 2024, ApJ, 974, 115

\bibitem[{Furuya {et~al.}(2022)Furuya, Hama, Oba, Kouchi, Watanabe, \&
  Aikawa}]{furuya_diffusion_2022}
Furuya, K., Hama, T., Oba, Y., {et~al.} 2022, ApJL, 933, L16

\bibitem[{{Furuya} {et~al.}(2022){Furuya}, {Tsukagoshi}, {Qi}, {Nomura},
  {Cleeves}, {Lee}, \& {Yoshida}}]{furuya_detection_2022}
{Furuya}, K., {Tsukagoshi}, T., {Qi}, C., {et~al.} 2022, \apj, 926, 148

\bibitem[{Grassi {et~al.}(2020)Grassi, Bovino, Caselli, Bovolenta, Vogt-Geisse,
  \& Ercolano}]{grassi_novel_2020}
Grassi, T., Bovino, S., Caselli, P., {et~al.} 2020, A\&A, 643, A155

\bibitem[{Grimme {et~al.}(2010)Grimme, Antony, Ehrlich, \&
  Krieg}]{grimme_consistent_2010}
Grimme, S., Antony, J., Ehrlich, S., \& Krieg, H. 2010, J. Chem. Phys., 132,
  154104

\bibitem[{Grimme {et~al.}(2011)Grimme, Ehrlich, \&
  Goerigk}]{grimme_effect_2011}
Grimme, S., Ehrlich, S., \& Goerigk, L. 2011, J. Comput. Chem., 32, 1456

\bibitem[{Groyne {et~al.}(2025)Groyne, Champagne, Baijot, \&
  Becker}]{Groyne_robust_2025}
Groyne, M., Champagne, B., Baijot, C., \& Becker, M.~D. 2025, A\&A, 698, A284

\bibitem[{Győrffy \& Werner(2018)}]{gyorffy_analytical_2018}
Győrffy, W. \& Werner, H.-J. 2018, J. Chem. Phys., 148, 114104

\bibitem[{Hama \& Watanabe(2013)}]{hama_surface_2013}
Hama, T. \& Watanabe, N. 2013, Chem. Rev., 113, 8783

\bibitem[{He {et~al.}(2016)He, Acharyya, \& Vidali}]{he_binding_2016}
He, J., Acharyya, K., \& Vidali, G. 2016, ApJ, 825, 89

\bibitem[{Helgaker {et~al.}(1997)Helgaker, Klopper, Koch, \&
  Noga}]{helgaker_basis-set_1997-2}
Helgaker, T., Klopper, W., Koch, H., \& Noga, J. 1997, J. Chem. Phys., 106,
  9639

\bibitem[{Herbst \& van Dishoeck(2009)}]{herbst_complex_2009}
Herbst, E. \& van Dishoeck, E.~F. 2009, ARA\&A, 47, 427

\bibitem[{Hjorth~Larsen {et~al.}(2017)Hjorth~Larsen, Jørgen~Mortensen,
  Blomqvist, Castelli, Christensen, Dułak, Friis, Groves, Hammer, Hargus,
  Hermes, Jennings, Bjerre~Jensen, Kermode, Kitchin, Leonhard~Kolsbjerg, Kubal,
  Kaasbjerg, Lysgaard, Bergmann~Maronsson, Maxson, Olsen, Pastewka, Peterson,
  Rostgaard, Schiøtz, Schütt, Strange, Thygesen, Vegge, Vilhelmsen, Walter,
  Zeng, \& Jacobsen}]{hjorth_larsen_atomic_2017}
Hjorth~Larsen, A., Jørgen~Mortensen, J., Blomqvist, J., {et~al.} 2017, J.
  Phys. Condens. Matter, 29, 273002

\bibitem[{Hunter(2007)}]{hunter_matplotlib_2007}
Hunter, J.~D. 2007, Comput. Sci. Eng., 9, 90

\bibitem[{Jeziorski {et~al.}(1994)Jeziorski, Moszynski, \&
  Szalewicz}]{jeziorski_perturbation_1994}
Jeziorski, B., Moszynski, R., \& Szalewicz, K. 1994, Chem. Rev., 94, 1887

\bibitem[{Kalvāns {et~al.}(2024)Kalvāns, Kalniņa, \&
  Veitners}]{kalvans_multi-grain_2024}
Kalvāns, J., Kalniņa, A., \& Veitners, K. 2024, A\&A, 687, A296

\bibitem[{Karssemeijer {et~al.}(2013)Karssemeijer, Ioppolo, Hemert, Avoird,
  Allodi, Blake, \& Cuppen}]{karssemeijer_dynamics_2013}
Karssemeijer, L.~J., Ioppolo, S., Hemert, M. C.~v., {et~al.} 2013, ApJ, 781, 16

\bibitem[{Keane {et~al.}(2001)Keane, Boogert, Tielens, Ehrenfreund, \&
  Schutte}]{keane_bands_2001}
Keane, J.~V., Boogert, A. C.~A., Tielens, A. G. G.~M., Ehrenfreund, P., \&
  Schutte, W.~A. 2001, A\&A, 375, L43

\bibitem[{Kruse \& Grimme(2012)}]{kruse_geometrical_2012}
Kruse, H. \& Grimme, S. 2012, J. Chem. Phys., 136, 154101

\bibitem[{Maggiolo {et~al.}(2019)Maggiolo, Gibbons, Cessateur, Keyser, Dhooghe,
  Gunell, Loreau, Mousis, \& Vaeck}]{maggiolo_effect_2019}
Maggiolo, R., Gibbons, A., Cessateur, G., {et~al.} 2019, ApJ, 882, 131

\bibitem[{Mariedahl {et~al.}(2018)Mariedahl, Perakis, Späh, Pathak, Kim,
  Camisasca, Schlesinger, Benmore, Pettersson, Nilsson, \&
  Amann-Winkel}]{mariedahl_x-ray_2018}
Mariedahl, D., Perakis, F., Späh, A., {et~al.} 2018, J. Phys. Chem. B, 122,
  7616

\bibitem[{McClure {et~al.}(2023)McClure, Rocha, Pontoppidan, Crouzet, Chu,
  Dartois, Lamberts, Noble, Pendleton, Perotti, Qasim, Rachid, Smith, Sun,
  Beck, Boogert, Brown, Caselli, Charnley, Cuppen, Dickinson, Drozdovskaya,
  Egami, Erkal, Fraser, Garrod, Harsono, Ioppolo, Jiménez-Serra, Jin,
  Jørgensen, Kristensen, Lis, McCoustra, McGuire, Melnick, Öberg, Palumbo,
  Shimonishi, Sturm, van Dishoeck, \& Linnartz}]{mcclure_ice_2023}
McClure, M.~K., Rocha, W. R.~M., Pontoppidan, K.~M., {et~al.} 2023, Nat Astron,
  7, 431

\bibitem[{Minissale {et~al.}(2022)Minissale, Aikawa, Bergin, Bertin, Brown,
  Cazaux, Charnley, Coutens, Cuppen, Guzman, Linnartz, McCoustra, Rimola,
  Schrauwen, Toubin, Ugliengo, Watanabe, Wakelam, \&
  Dulieu}]{minissale_thermal_2022}
Minissale, M., Aikawa, Y., Bergin, E., {et~al.} 2022, ACS Earth Space Chem., 6,
  597

\bibitem[{Molpeceres {et~al.}(2024)Molpeceres, Furuya, \&
  Aikawa}]{molpeceres_enhanced_2024}
Molpeceres, G., Furuya, K., \& Aikawa, Y. 2024, A\&A, 688, A150

\bibitem[{Molpeceres \& Kästner(2020)}]{molpeceres_adsorption_2020}
Molpeceres, G. \& Kästner, J. 2020, PCCP, 22, 7552

\bibitem[{Molpeceres {et~al.}(2023)Molpeceres, Zaverkin, Furuya, Aikawa, \&
  Kästner}]{molpeceres_reaction_2023}
Molpeceres, G., Zaverkin, V., Furuya, K., Aikawa, Y., \& Kästner, J. 2023,
  A\&A, 673, A51

\bibitem[{Molpeceres {et~al.}(2020)Molpeceres, Zaverkin, \&
  Kästner}]{molpeceres_neural-network_2020}
Molpeceres, G., Zaverkin, V., \& Kästner, J. 2020, MNRAS, 499, 1373

\bibitem[{{Muenter}(1975)}]{Muenter1975}
{Muenter}, J.~S. 1975, J. Mol. Spectrosc., 55, 490

\bibitem[{Nagasawa {et~al.}(2021)Nagasawa, Sato, Hasegawa, Numadate, Shioya,
  Shimoaka, Hasegawa, \& Hama}]{nagasawa_absolute_2021}
Nagasawa, T., Sato, R., Hasegawa, T., {et~al.} 2021, ApJL, 923, L3

\bibitem[{Neese {et~al.}(2020)Neese, Wennmohs, Becker, \&
  Riplinger}]{neese_orca_2020}
Neese, F., Wennmohs, F., Becker, U., \& Riplinger, C. 2020, J. Chem. Phys.,
  152, 224108

\bibitem[{Nguyen {et~al.}(2018)Nguyen, Baouche, Congiu, Diana, Pagani, \&
  Dulieu}]{nguyen_segregation_2018}
Nguyen, T., Baouche, S., Congiu, E., {et~al.} 2018, A\&A, 619, A111

\bibitem[{Nguyen {et~al.}(2020)Nguyen, Oba, Shimonishi, Kouchi, \&
  Watanabe}]{nguyen_experimental_2020}
Nguyen, T., Oba, Y., Shimonishi, T., Kouchi, A., \& Watanabe, N. 2020, ApJL,
  898, L52

\bibitem[{Noble {et~al.}(2012{\natexlab{a}})Noble, Dulieu, Congiu, \&
  Fraser}]{noble_1_2012}
Noble, J., Dulieu, F., Congiu, E., \& Fraser, H. 2012{\natexlab{a}}, EAS Publ.
  Ser., 58, 353

\bibitem[{Noble {et~al.}(2012{\natexlab{b}})Noble, Congiu, Dulieu, \&
  Fraser}]{noble_thermal_2012}
Noble, J.~A., Congiu, E., Dulieu, F., \& Fraser, H.~J. 2012{\natexlab{b}},
  MNRAS, 421, 768

\bibitem[{Noble {et~al.}(2024)Noble, Fraser, Smith, Dartois, Boogert, Cuppen,
  Dickinson, Dulieu, Egami, Erkal, Giuliano, Husquinet, Lamberts, Maté,
  McClure, Palumbo, Shimonishi, Sun, Bergner, Brown, Caselli, Congiu,
  Drozdovskaya, Herrero, Ioppolo, Jimenez-Serra, Linnartz, Melnick, McGuire,
  Oberg, Perotti, Qasim, Rocha, \& Urso}]{noble_detection_2024}
Noble, J.~A., Fraser, H.~J., Smith, Z.~L., {et~al.} 2024, Nat Astron, 8, 1169

\bibitem[{Notsu(2020)}]{notsu_water_2020}
Notsu, S. 2020, Water {Snowline} in {Protoplanetary} {Disks}, Springer {Theses}
  (Singapore)

\bibitem[{Papajak {et~al.}(2011)Papajak, Zheng, Xu, Leverentz, \&
  Truhlar}]{papajak_perspectives_2011}
Papajak, E., Zheng, J., Xu, X., Leverentz, H.~R., \& Truhlar, D.~G. 2011, J.
  Chem. Theory Comput., 7, 3027

\bibitem[{Parker {et~al.}(2014)Parker, Burns, Parrish, Ryno, \&
  Sherrill}]{parker_levels_2014}
Parker, T.~M., Burns, L.~A., Parrish, R.~M., Ryno, A.~G., \& Sherrill, C.~D.
  2014, J. Chem. Phys., 140, 094106

\bibitem[{Perrero {et~al.}(2022)Perrero, Enrique-Romero, Ferrero, Ceccarelli,
  Podio, Codella, Rimola, \& Ugliengo}]{perrero_binding_2022-1}
Perrero, J., Enrique-Romero, J., Ferrero, S., {et~al.} 2022, ApJ, 938, 158

\bibitem[{Poštulka {et~al.}(2025)Poštulka, Slavíček, Kästner, \&
  Molpeceres}]{postulka_diffusive_2025}
Poštulka, J., Slavíček, P., Kästner, J., \& Molpeceres, G. 2025, A\&A, 697,
  A51

\bibitem[{Raghavachari {et~al.}(1989)Raghavachari, Trucks, Pople, \&
  Head-Gordon}]{raghavachari_fifth-order_1989}
Raghavachari, K., Trucks, G.~W., Pople, J.~A., \& Head-Gordon, M. 1989, Chem.
  Phys. Lett., 157, 479

\bibitem[{Sameera {et~al.}(2021)Sameera, Senevirathne, Andersson, Al-lbadi,
  Hidaka, Kouchi, Nyman, \& Watanabe}]{sameera_ch3o_2021}
Sameera, W. M.~C., Senevirathne, B., Andersson, S., {et~al.} 2021, J. Phys.
  Chem. A, 125, 387

\bibitem[{Schriber {et~al.}(2021)Schriber, Sirianni, Smith, Burns, Sitkoff,
  Cheney, \& Sherrill}]{schriber_optimized_2021}
Schriber, J.~B., Sirianni, D.~A., Smith, D. G.~A., {et~al.} 2021, J. Chem.
  Phys., 154, 234107

\bibitem[{Seung {et~al.}(1992)Seung, Opper, \& Sompolinsky}]{seung_query_1992}
Seung, H.~S., Opper, M., \& Sompolinsky, H. 1992, in Proceedings of the fifth
  annual workshop on {Computational} learning theory, Pittsburgh Pennsylvania
  USA, 287--294

\bibitem[{Simons {et~al.}(2020)Simons, Lamberts, \&
  Cuppen}]{simons_formation_2020}
Simons, M. a.~J., Lamberts, T., \& Cuppen, H.~M. 2020, A\&A, 634, A52

\bibitem[{Smith {et~al.}(2016)Smith, May, \& Kay}]{smith_desorption_2016}
Smith, R.~S., May, R.~A., \& Kay, B.~D. 2016, J. Phys. Chem. B, 120, 1979

\bibitem[{Sylvetsky {et~al.}(2017)Sylvetsky, Kesharwani, \&
  Martin}]{sylvetsky_aug-cc-pvnz-f12_2017}
Sylvetsky, N., Kesharwani, M.~K., \& Martin, J. M.~L. 2017, J. Chem. Phys.,
  147, 134106

\bibitem[{Szalewicz(2012)}]{szalewicz_symmetry-adapted_2012}
Szalewicz, K. 2012, WIREs Comput. Mol. Sci., 2, 254

\bibitem[{Tinacci {et~al.}(2023)Tinacci, Germain, Pantaleone, Ceccarelli,
  Balucani, \& Ugliengo}]{tinacci_theoretical_2023}
Tinacci, L., Germain, A., Pantaleone, S., {et~al.} 2023, ApJ, 951, 32

\bibitem[{Tinacci {et~al.}(2022)Tinacci, Germain, Pantaleone, Ferrero,
  Ceccarelli, \& Ugliengo}]{tinacci_theoretical_2022}
Tinacci, L., Germain, A., Pantaleone, S., {et~al.} 2022, ACS Earth and Space
  Chemistry, 6, 1514

\bibitem[{Wakelam {et~al.}(2017)Wakelam, Loison, Mereau, \&
  Ruaud}]{wakelam_binding_2017}
Wakelam, V., Loison, J.~C., Mereau, R., \& Ruaud, M. 2017, Mol. Astrophys., 6,
  22

\bibitem[{Wang(2023)}]{wang_leepinggeometric_2023}
Wang, L.-P. 2023, leeping/{geomeTRIC}

\bibitem[{Watanabe \& Kouchi(2002)}]{Watanabe2002}
Watanabe, N. \& Kouchi, A. 2002, ApJ, 571, L173

\bibitem[{Weigend \& Ahlrichs(2005)}]{weigend_balanced_2005}
Weigend, F. \& Ahlrichs, R. 2005, PCCP, 7, 3297

\bibitem[{Werner {et~al.}(2012)Werner, Knowles, Knizia, Manby, \&
  Schütz}]{werner_molpro_2012}
Werner, H.-J., Knowles, P.~J., Knizia, G., Manby, F.~R., \& Schütz, M. 2012,
  WIREs Comput. Mol. Sci., 2, 242

\bibitem[{Werner {et~al.}(2020)Werner, Knowles, Manby, Black, Doll, Heßelmann,
  Kats, Köhn, Korona, Kreplin, Ma, Miller, Mitrushchenkov, Peterson, Polyak,
  Rauhut, \& Sibaev}]{werner_molpro_2020}
Werner, H.-J., Knowles, P.~J., Manby, F.~R., {et~al.} 2020, J. Chem. Phys.,
  152, 144107

\bibitem[{Zaverkin {et~al.}(2022)Zaverkin, Holzmüller, Schuldt, \&
  Kästner}]{Zaverkin2022}
Zaverkin, V., Holzmüller, D., Schuldt, R., \& Kästner, J. 2022, J. Chem.
  Phys., 156, 114103

\bibitem[{Zaverkin {et~al.}(2021)Zaverkin, Holzmüller, Steinwart, \&
  Kästner}]{zaverkin_fast_2021}
Zaverkin, V., Holzmüller, D., Steinwart, I., \& Kästner, J. 2021, J. Chem.
  Theory Comput., 17, 6658

\bibitem[{Zaverkin \& Kästner(2020)}]{zaverkin_gaussian_2020}
Zaverkin, V. \& Kästner, J. 2020, J. Chem. Theory Comput., 16, 5410

\bibitem[{Zhao \& Truhlar(2005)}]{Zhao2005}
Zhao, Y. \& Truhlar, D.~G. 2005, J. Phys. Chem. A, 109, 5656

\end{thebibliography}

\begin{appendix}

\section{Density functional theory benchmark} \label{sec:ap_DFTbench}
\subsection{Reference energy}
We used a total of five different binding sites to obtain 
a high-level binding energy form wavefunction methods.  We 
optimized the binding site geometry and the CCSD(T)-F12/cc-pVDZ-F12
We computed the  BEs on the different binding sites 
using the focal point analysis (FPA) technique to obtain a CCSD(T)/CBS quality value.  
In this approach, energy values  computed at the SCF, MP2, CCSD, CCSD(T), 
levels of theory with the aug-cc-pVXZ basis sets. The total  energy was extrapolated 
to the complete basis set limit (CBS) using  the functional form:
\[
E_{\text{SCF}}(X) = A + \left( B \, e^{-CX} \right)
\]
\[
E_{\text{CORR}} = E + FX^{-3}
\]
where $E_{\text{SCF}}$ and $E_{\text{CORR}}$ are the extrapolated SCF and 
correlation energies (MP2, CCSD, CCSD(T)), respectively, and $X$ is the cardinal number corresponding 
to the maximum angular momentum of the aug-cc-pVXC  basis set (X=D,T,Q). 
$A$, $B$, $C$, $E$, and $F$  are fitting parameters for the SCF and correlation energies, respectively.
We used the extrapolation libraries implemented in the Psi4 driver together
with the BEEP energy benchmark workflow. The binding energy was computed as usual:
The binding energy (BE) of a species ($i$) adsorbed on a surface ($ice$) is defined as:    
\begin{equation}
BE(i) = E_\mathrm{sup} - (E_\mathrm{W}+ E_i)
\label{eq:BE}
\end{equation}
where $E_\mathrm{sup}$ is the energy of the super-molecule (CO+W) formed by the adsorbate bound 
to the water cluster, $E_\mathrm{W}$ refers to the water cluster energy, and $E_i$ is the energy of the adsorbate. 
Tables \ref{tab:fp_COdia}–\ref{tab:fp_COtrc} show the evolution of the contribution 
of electron correlation to the binding energy. In general the correlation converges smoothly 
when increasing the correlation level and basis set size, which gives binding energies 
of sub-chemical accuracy (uncertainty of approximately 0.1\kcal $\approx$ 50 K) so any DFT functional 
which yields BEs close to that accuracy is a good choice of model chemistry. Table \ref{tab:IE_DE_BE}
shows the CCSD(T)/CBS values for the IE, DE, and BE of each of the model systems.

\begin{table}[H]
\centering
\caption{Focal point analysis of the binding energy for the \COdia\ structure.}
\label{tab:fp_COdia}
\begin{tabular*}{0.49\textwidth}{@{\extracolsep{\fill}}lccccc}
\hline
                & BE$_{\text{SCF}}$ & $+\delta$MP2  & $+\delta$CCSD   & $+\delta$CCSD(T)  & NET \\ 
\hline
aDZ     & 0.96  & 2.49  & -0.50  & 0.40  & [3.34]  \\ 
aTZ     & 0.83  & 2.67  & -0.54  & 0.41  & [3.37]  \\ 
aQZ     & 0.70  & 2.61  & --     & --    & [3.31]  \\ 
CBS     & 0.61  & 2.56  & -0.56  & 0.42  & \textbf{[3.03]} \\ 
\hline
\end{tabular*}
\tablefoot{Final binding energy: BE[CCSD(T)/CBS] = 3.03 \kcal (1522 K).}
\end{table}

\begin{table}[H]
\centering
\caption{Focal point analysis in \kcal of the binding energy for the \COdib\ structure.}
\label{tab:fp_COdib}
\begin{tabular*}{0.49\textwidth}{@{\extracolsep{\fill}}lccccc}
\hline
                & BE$_{\text{SCF}}$ & $+\delta$MP2  & $+\delta$CCSD   & $+\delta$CCSD(T)  & NET \\ 
\hline
aDZ     & 1.20  & 0.85  & 0.10  & 0.26  & [2.40]  \\ 
aTZ     & 1.04  & 0.92  & 0.09  & 0.24  & [2.29]  \\ 
aQZ     & 0.95  & 0.86  & --    & --    & [1.81]  \\ 
CBS     & 0.89  & 0.82  & 0.09  & 0.23  & \textbf{[2.02]} \\ 
\hline
\end{tabular*}
\tablefoot{Final binding energy: BE[CCSD(T)/CBS] = 2.02 kcal (1017 K).}
\end{table}

\begin{table}[H]
\centering
\caption{Focal point analysis in \kcal of the binding energy for the \COtra\ structure.}
\label{tab:fp_COtra}
\begin{tabular*}{0.49\textwidth}{@{\extracolsep{\fill}}lccccc}
\hline
                & BE$_{\text{SCF}}$ & $+\delta$MP2  & $+\delta$CCSD   & $+\delta$CCSD(T)  & NET \\ 
\hline
aDZ     & 0.47  & 1.72  & -0.25  & 0.29  & [2.24]  \\ 
aTZ     & 0.22  & 1.81  & -0.28  & 0.29  & [2.04]  \\ 
aQZ     & 0.14  & 1.74  & --     & --    & [1.88]  \\ 
CBS     & 0.10  & 1.69  & -0.30  & 0.29  & \textbf{[1.78]} \\ 
\hline
\end{tabular*}
\tablefoot{Final binding energy: BE[CCSD(T)/CBS] = 1.78 \kcal (896 K).}
\end{table}

\begin{table}[H]
\centering
\caption{Focal point analysis in \kcal of the binding energy for the \COtrb structure.}
\label{tab:fp_COtrb}
\begin{tabular*}{0.49\textwidth}{@{\extracolsep{\fill}}lccccc}
\hline
                & BE$_{\text{SCF}}$ & $+\delta$MP2  & $+\delta$CCSD   & $+\delta$CCSD(T)  & NET \\ 
\hline
aDZ     & 0.43  & 1.54  & -0.07  & 0.32  & [2.22]  \\ 
aTZ     & 0.18  & 1.55  & -0.10  & 0.30  & [1.94]  \\ 
aQZ     & 0.09  & 1.48  & --     & --    & [1.56]  \\ 
CBS     & 0.03  & 1.43  & -0.11  & 0.29  & \textbf{[1.64]}\\ 
\hline
\end{tabular*}
\tablefoot{Final binding energy: BE[CCSD(T)/CBS] = 1.64 \kcal (826 K).}
\end{table}

\begin{table}[H]
\centering
\caption{Focal point analysis in \kcal of the binding energy for the \COtrc\ structure.}
\label{tab:fp_COtrc}
\begin{tabular*}{0.49\textwidth}{@{\extracolsep{\fill}}lccccc}
\hline
                & BE$_{\text{SCF}}$ & $+\delta$MP2  & $+\delta$CCSD   & $+\delta$CCSD(T)  & NET \\ 
\hline
aDZ     & 0.12  & 1.40  & -0.29  & 0.25  & [1.23]  \\ 
aTZ     & 0.14  & 1.36  & -0.31  & 0.22  & [1.14]  \\ 
aQZ     & 0.29  & 1.24  & --     & --    & [0.96]  \\ 
CBS     & 0.43  & 1.16  & -0.31  & 0.21  & \textbf{[0.63]} \\ 
\hline
\end{tabular*}
\tablefoot{Final binding energy: BE[CCSD(T)/CBS] = 0.63 \kcal (316 K).}
\end{table}

\begin{table}[H]
\centering
\caption{Interaction energy, deformation energy, and binding energy at the CCSD(T)/CBS level of theory.}
\label{tab:IE_DE_BE}
\begin{tabular*}{0.49\textwidth}{@{\extracolsep{\fill}}lccc}

\hline
Sample & IE & DE & BE \\
\hline
\COdia & 1647 & -124  & 1522  \\
\COdib & 1097 & -79   & 1017  \\
\COtra & 960  & -64   & 896   \\
\COtrb & 880  & -54   & 826   \\
\COtrc & 1907 & -1591 & 316   \\
\hline
\end{tabular*}
\end{table}

\begin{table}[H]
    \centering
    \caption{Interaction energy decomposition at the SAPT2+/aug-cc-PVDZ level of theory}
    \begin{tabular*}{0.49\textwidth}{@{\extracolsep{\fill}}lccccccc}
    \hline
    Structure &  $E_\mathrm{elst}$ & $E_\mathrm{exch}$ &  $E_\mathrm{ind}$ &  $E_\mathrm{disp}$ &   \textbf{IE} \\  
    \hline
         \COdia &  2405 & -2729 &  773 &   1202 &  1651 \\
         \COdib & 1008  & -1569 & 451 &   951 &   840  \\
\COtra & 1180&  -1265 & 214 &  857 & 985 \\
\COtrb & 707 & -1149  &163  & 1004 & 726 \\
\COtrc & 3101 & -3958  &1170  & 1539 & 1852\\
 \hline
    \end{tabular*}
   
    \label{tab:sapt_ref}
\end{table}
Table \ref{tab:sapt_ref} shows the different contributions to the IE, estimated at SAPT2+/aug-CC-PVDZ level.
For the strongest binding mode in which the H-bond is established through
the carbon end of the CO molecule.  In both cases, the electrostatic energy represents 
the largest component of the interaction energy. Nonetheless, the electrostatic 
interaction energy in the dimer configuration 
is more than twofold that of the trimer. 
The enhanced interaction of CO within the water dimer
is attributable to its engagement with a water molecule that solely acts as a H-bond
acceptor, thereby rendering the water molecule more electron deficient and thus 
intensifying the electrostatic coupling with the negatively charged carbon end of 
the CO molecule. This is not the case in the trimer structure in which all water molecules 
are both receiving and donating H-bonds within a H-bond cycle. Furthermore this difference 
is accentuated in the polarization interaction, as the water dimer has a much greater ability
to polarize the CO  molecule than the water trimer. The dispersion interaction is significantly
lower than the electrostatic interaction in both molecules, however for the trimer the dispersion 
represents a larger percentage of the total interaction energy with a dispersion factor
of 0.3 compared to the water dimer structure which has a dispersion factor of 0.21. Finally 
due to the closer proximity of the CO molecule to the water dimer in its bound conformation,
the exchange repulsion is quite significant and higher than in the water trimer.  In summary,
CO binds to the small water clusters predominately through electrostatic interactions. This
contribution is larger when CO is directly interacting  with a dangling water molecule as 
in the water dimer. However whenever a dangling-H bond is present, the interaction is weaker
but the predominant mode of interaction is still electrostatic.

\subsection{DFT geometry and binding energy benchmark result}
We  probed 6 hybrid and meta-hybrid GGA functionals against the CCSD(T)-F12/cc-PVDZ 
reference geometry to evaluate which functional can best reproduce the binding 
site geometry of the \COdib and \COtrb structures. 
The MPWB1K-D3BJ functional displays the lowest mean RMSD 
for describing the binding sites geometry of the model systems.
Furthermore, we conducted a BE benchmark study on MPWB1K-D3BK/def2-TZVP 
geometries. The structures used for this benchmark
were \COdia, \COdib and \COtrb. The energies were compared to 
the BE values listed in Tables \ref{tab:fp_COdia}, Table \ref{tab:fp_COdib} and Table
\ref{tab:fp_COtrb}. The results of the best 15  
of a total of 166 hybrid, meta-hybrid and long-range corrected DFT functionals
MAE are shown in Table B.2. Again MPWB1K-D3BJ is one 
of the best performing DFT functionals with an average MAE of 63 K, 
which is within  40 K of the best performing functional.  
Given the good  combined performance in gradients and BEs, 
the MPWB1K-D3BJ/def2-TZVP//MPWB1K-D3BJ/def2-TZVP model chemistry proves to 
be a suitable option for MLP training of gradients and energies. 
Nonetheless,  it is important  to note that this level of theory carries an 
uncertainty of at least 63 K in any BE prediction.

\begin{table}[h!]
\centering
\caption{RMSD for the binding site geometry of the \COdib and \COtrb
structure with different hybrid and meta\-hybrid DFT functionals.  }

\begin{tabular*}{0.49\textwidth}{@{\extracolsep{\fill}}lc}
\hline
Level of theory         & RMSD \\
\hline
M05-2X/def2-TZVP                & 0.034 \\
MPWB1K-D3BJ/def2-TZVP           & 0.031 \\
PWB6K-D3BJ/def2-TZVP            & 0.031 \\
B3LYP-D3BJ/def2-TZVP            & 0.043 \\
B3PW91-D3BJ/def2-TZVP           & 0.053 \\
PBE0-D3BJ/def2-TZVP            & 0.047 \\
\hline
HF-3c/MINIX                      & 0.190 \\
PBEh-3c/def2-mSVP               & 0.106 \\
\hline
\end{tabular*}
\end{table}

\begin{table}[h!]
\centering
\caption{Mean absolute error (MAE) in K of binding energies for benchmarked DFT functionals.}

\begin{tabular*}{0.49\textwidth}{@{\extracolsep{\fill}}lc}
\hline
Level of theory &  MAE (K)  \\
\hline
MN12-SX-D3BJ/def2-TZVP & 21 \\
LC-VV10/def2-TZVP      & 24 \\
$\omega$-PBE-D3MBJ/def2-TZVP   & 39 \\
$\omega$-B97M-V/def2-TZVP      & 44 \\
PW6B95-D3BJ/def2-TZVP  & 45 \\
M06-2X/def2-TZVP       & 47 \\
REVPBE0-D3BJ/def2-TZVP & 50 \\
M05/def2-TZVP          & 51 \\
REVPBE0-NL/def2-TZVP   & 59 \\
MPWB1K-D3BJ/def2-TZVP  & 63 \\
$\omega$-B97M-D3BJ/def2-TZVP   & 65 \\
B97-2-D3BJ/def2-TZVP   & 70 \\
M11-D3BJ/def2-TZVP     & 75 \\
M08-HX/def2-TZVP       & 76 \\
B3LYP-D3BJ/def2-TZVP & 141 \\
\bottomrule
\end{tabular*}
\tablefoot{
 The reference energy is CCSD(T)/CBS. The table includes some of the best and popular out of a total of 166 meta-hybrid, hybrid, and long-range corrected XC functionals. The structures included in the binding energy benchmark are: \COdia, \COdib and  \COtrb }
\end{table}

\section{ZPVE correction for the BEs}
\label{sec:ap_zpve}

We determined ZPVE correction at the CCSD(T)-F12/cc-pVDZ-F12 level of theory through the calculation of the Hessian matrix and subsequent harmonic vibrational analysis on the four binding sites used in the DFT benchmark.  Furthermore, we re-optimized the structures using DFT, specifically MPWB1K-D3BJ / def2-TZVP, and calculated the ZPVE correction at this level of theory as well. The ZPVE correction is reported as:

\begin{equation}
    \Delta_{\text{ZPVE}} = \text{ZPVE}(W_X + CO) - \text{ZPVE}(W_X) + \text{ZPVE}(W_X)
\end{equation}

where $X = 2,3$. 
We also analyzed the impact of anharmonic correction
to the vibrational frequencies using vibrational
perturbation theory at the CCSD(T)/aug-cc-pVDZ level of theory using 
CFOUR software.  We applied these corrections to two model systems, 
namely \COdia and \COtra. The results are shown in Table \ref{zpve_anharm}. The anharmonic 
correction results to be  small and lies within the error of the DFT method. 
Nonetheless in both cases it lowers the ZPVE correction and therefore we applied the 
anharmonic correction of $-$0.05 \kcal to all the ZPVE corrections.
The ZPVE  correction that we used to calculate the scaling factor 
that we applied to all the binding energies reported in this 
work is based on the ZPVE corrected BEs at the CCSD(T)-F12/cc-pVDZ 
level of theory:

\begin{equation}
    BE_\mathrm{ZPVE} =  \text{BE(CCSD(T)/cc-pVDZ)} + \Delta_{\text{ZPVE}} + \Delta_{\text{anharm}}
\end{equation}

The result for ZPVE correction in the different model systems and the scaling factors is shown in Table \ref{tab:zpve_corr}.
The scaling factor for the BEs is calculated as:
\begin{equation}
    \text{ZPVE}_{SF} = \frac{BE_{\text{ZPVE}}}{BE}
\end{equation}
As a result of this analysis, we scaled all the BEs obtained by 
our MLP by 0.677 to account for ZPVE. The ZPVE correction is 
inherently a local phenomenon, as the difference primarily 
arises from the intermolecular vibration of the bound CO 
relative to the water molecules. This approach is likely more 
accurate than computing vibrational frequencies on the ASW, 
where spurious imaginary frequencies in the hydrogen bond 
network can distort the evaluation of the ZPVE.

\begin{table}[h!]
\centering
\caption{Zero-point vibrational energies and anharmonic corrections in kelvin}
\label{zpve_anharm}
\begin{tabular*}{0.49\textwidth}{@{\extracolsep{\fill}}lccc}
\hline
Structure  & ZPVE [K] & ZPVE$_\text{anharm}$ [K] & \textbf{$\Delta_{\text{anharm}}$ [K]} \\
\hline
\COdib        & $-$383              & $-$358                      & 27                                \\
\COtrb        & $-$228              & $-$201                     & 27                                \\
\hline
\end{tabular*}
\end{table}

\begin{table}[h!]
\centering
\caption{Binding energy and ZPVE values obtained at the CCSD(T)-F12/cc-pVDZ level of theory. 
}
\label{tab:zpve_corr}
\begin{tabular*}{0.49\textwidth}{@{\extracolsep{\fill}}lcccc}

\hline
Structure & BE  & $\Delta$ ZPVE   & BE+ZPVE + $\Delta_\mathrm{anh}$ & \textbf{ZPVE \%} \\
\hline
\COdib   &  999 & -358   & 641    & 0.641            \\
\COdia   &  1483 & -567   & 917   & 0.618           \\
\COtra   &  874 & -259   & 616    & 0.704            \\
\COtrb   &  825 & -201   & 624    & 0.756            \\
\COtrc   &  352 & -118   & 234    & 0.665          \\
\hline
\textbf{Average} & 905  & -311  & 604  & 0.677            \\
\hline
\end{tabular*}
\tablefoot{
Then anharmonic (anh) corrections are computed at the CCSD(T)/cc-pVDZ level of theory.
}
\end{table}

\section{MLP training and validation}
\label{sec:ap_MLP}

\subsection{Training set composition}
\label{sec:ap_trainingset}

Table \ref{tab:data_set} shows the composition of the training and test set for the MLP we used in this study.
The training set configurations consist of 8321
structures, generated via MD simulations at various temperatures (100, 300 K) on water
clusters of different sizes as well as equilibrium
structures where the CO molecule interact with the water clusters. 
 The propagation step, used for sampling geometries, was carried out employing GFN2-xTB \citep{bannwarth_gfn2-xtbaccurate_2019}, while energy and gradients are computed at
MPWB1K-D3BJ/def2-TZVP level of theory.

\begin{table*}[h]
\small
    \centering
    \caption{Training set composition and test set accuracy for the machine learning potential.}
\begin{tabular*}{\textwidth}{@{\extracolsep{\fill}}lcccccc}
\hline
        Type & Formula & Training (66\%)  & \multicolumn{2}{c}{Energies [meV/atom]} & \multicolumn{2}{c}{Forces [meV/\text{\AA}]} \\
& & &   MAE  & RMSE & MAE & RMSE \\
\hline
\ce{W32} &  MD  & 820 &  0.318 & 0.416 & 20.284 & 28.484 \\
\ce{W70} &  MD & 122 & 0.358 & 0.426 & 28.793 & 39.812 \\
\ce{W49} &  MD & 414  & 0.252 & 0.304 & 24.148 & 33.663 \\
\ce{W22-CO} & MD / metadynamics  & 3090 & 0.390 & 0.508 & 22.972 & 32.734 \\
\ce{W37-CO} & MD / metadynamics & 2195 & 0.322 & 0.401 & 24.762 & 34.070 \\
\ce{W50-CO}  &   active learning & 1540 & 0.589 & 0.770 & 24.348 & 33.548 \\
\ce{W60-CO} & active learning  & 140 & 0.268 & 0.319 & 27.271 & 36.718 \\

\hline
    TOTAL & & 8321  &  0.393 & 0.524 & 23.970 & 33.384 \\
\hline
    \end{tabular*}
    \label{tab:data_set}
    \tablefoot{The training set comprised 66\% of the total data, while the test set and validation set each contained 16\%.}
\end{table*}

\subsection{Test set performance}
\begin{figure*}
    \centering
    \includegraphics[width=0.9\linewidth]{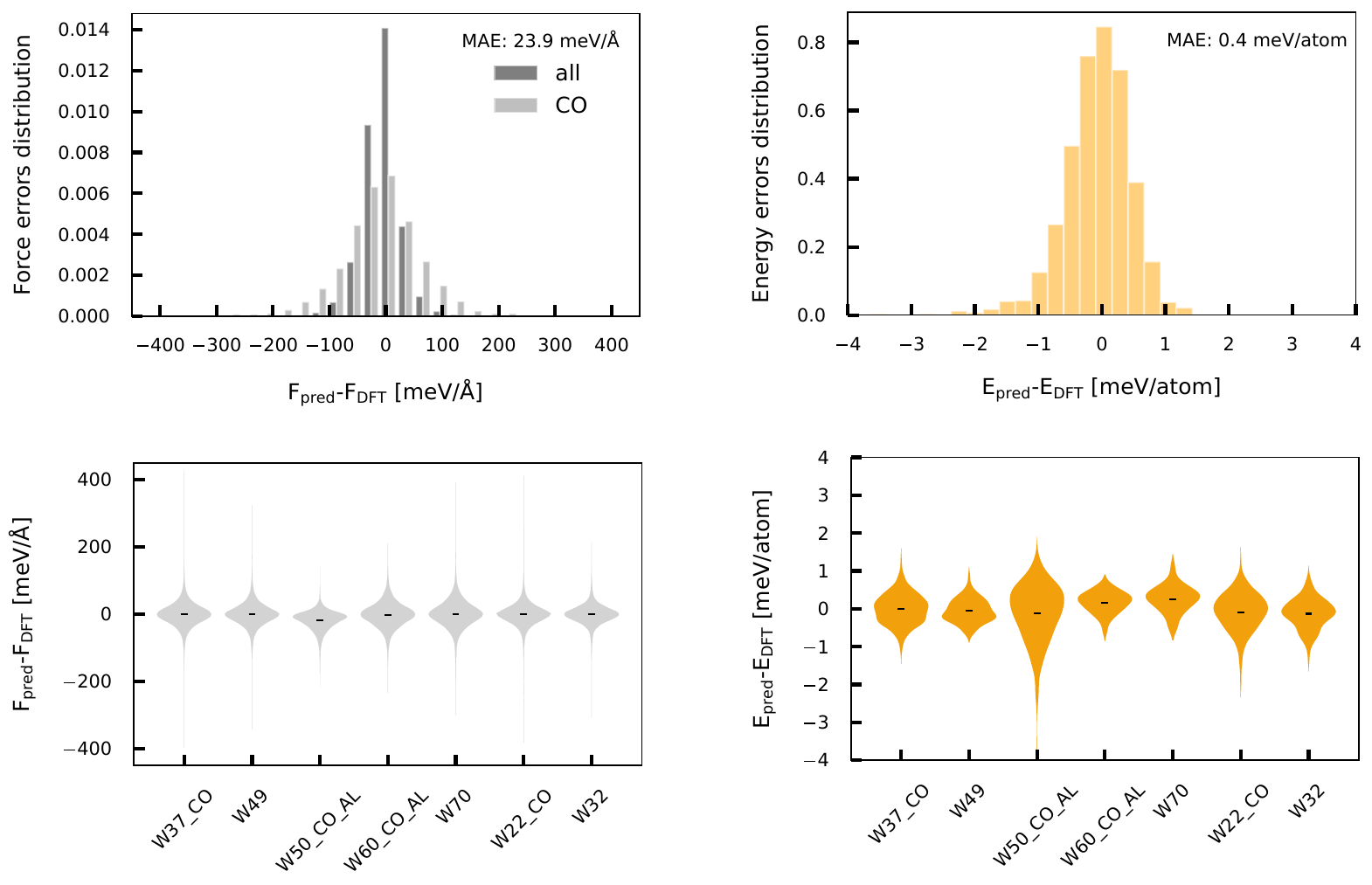} \\
    \caption{
Force and energy prediction errors for the test set (16\% of the dataset) are summarized across different molecular systems. The upper plots highlight overall errors, while in  the lower section, violin plots show the distribution of errors in both forces (left) and energies (right), grouped by chemical formula. Mean errors for each system are marked to aid comparison.
}
    \label{fig:error_mlp}
\end{figure*}

The test set used to evaluate the fidelity of the trained MLP,
mirrors the composition of the training set as detailed
in Table \ref{tab:data_set} and consists of 1660 structures. The results, 
as illustrated in the Fig. \ref{fig:error_mlp}, show a MAE of 23.9 meV/$\text{\AA}$
for forces and 0.4 meV/atom for energies across all structures, 
indicating good overall accuracy in predicting both forces and energies. Histograms of the force and 
energy errors reveal a tight distribution centered around zero, with minimal outliers, 
further underscoring the reliability of the model. Fig. \ref{fig:error_mlp}  bottom panel depicts the 
individual performance of the subsets within the test set. The performance is consistent across 
different cluster sizes and compositions, with deviations typically within acceptable 
ranges. The test subset containing the structures from the active learning round (W50\_CO\_Al) 
exhibits the highest deviation, which is expected given its significant 
morphological variations; this set includes CO bound to water clusters derived from the ASW$_{500}$
porous and non-porous surfaces, which exhibit a variety of ASW like environments 
and are therefore more difficult to reproduce. However the MAE of this group is still 
within an acceptable range of uncertainty in order to produce reliable binding energies. 

\subsection{Comparison with DFT binding energies}

Table \ref{tab:orca_valid} shows the mean values of BEs 
computed on a test set  of 22 binding sites of CO bound to a water cluster 
of 22  water molecules. The BEs are computed with the MLP used in this work
and at the  MPWB1K-D3BJ level of theory. There is an excellent agreement
in both the BE and the standard deviation for this set with a deviation of 
only 16 K, confirming the 
ability of the MLP to compute BEs for the CO + ice system. 

\begin{table}[h!]
\centering
\caption{Comparison of mean binding energies for CO-W$_{22}$ binding sites.}
\begin{tabular*}{0.49\textwidth}{@{\extracolsep{\fill}}lcc}
\hline
  & MLP & DFT  \\
\hline
BE            &  947              &  963                                                \\
Std Dev        & 276              &  260                                    \\
\hline
\end{tabular*}
\label{tab:orca_valid}
\tablefoot{The comparison includes 22 distinct binding sites. Geometries were optimized and binding energies (BE) were computed at the MPWB1K-D3BJ/def2-TZVP level of theory. All values are in K.}
\end{table}

\subsection{Validation across machine learning models}

\begin{figure*}
    \centering
    \includegraphics[width=0.9\linewidth]{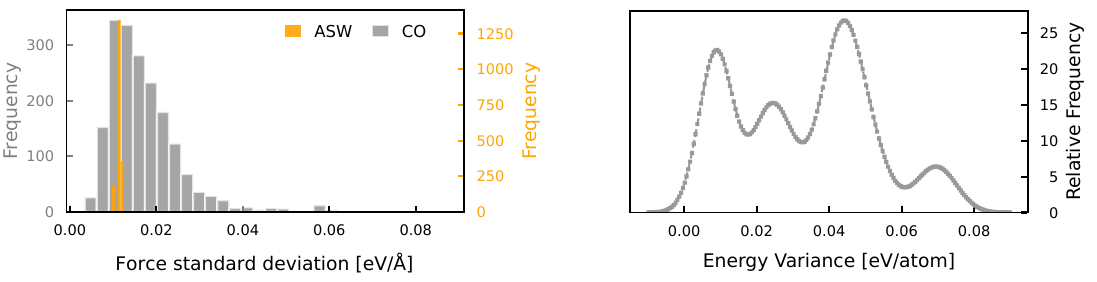} \\
    \caption{Distribution of standard deviations of 
    predicted forces (left panel) and energy variances 
    (right panel) for all the binding sites of CO on ASW (\textit{np}ASW and \textit{p}ASW), obtained from three independently trained machine learning potentials derived from the same training dataset.}
    \label{fig:model_dev}
\end{figure*}

To assess the robustness and reliability of the MLPs, we
examined the standard deviations of predicted forces and 
energy variances from three independently trained ML models 
for CO adsorbed on ASW, considering all binding sites on \textit{np}ASW and \textit{p}ASW. The results are shown in Fig.~\ref{fig:model_dev}.  
The left panel of Fig.~\ref{fig:model_dev} shows the frequency 
distributions of the standard deviations of predicted forces 
(in eV/\AA). Both CO molecules and ASW atoms exhibit force 
standard deviations mostly below 0.02 eV/\AA. The 
higher variability observed for CO molecules compared to ASW 
atoms can be attributed to the broader chemical space spanned 
by CO when bound to the ASW surfaces, relative to the training 
set, resulting in larger uncertainties in predictions. 
Nevertheless, the overall error in force predictions for CO 
remains small, confirming the reliability of the MLPs.
In the right panel of Fig. \ref{fig:model_dev}, the relative frequency 
distribution of the energy variance (in eV/atom) is shown. The 
multimodal distribution indicates distinct groups of 
configurations, reflecting variations in energy variance
across different types of ice surfaces. Despite these 
variations, the overall low magnitude of energy variances
underscores the suitability and reliability of these MLPs
for accurately predicting binding energies of CO on ASW.
The disagreement between the models in
predicted quantities, atomic forces, in this case, is calculated as:
\begin{equation}
\sigma = \sqrt{\frac{1}{3 N_{\text{models}}} \sum_{i=1}^{N_{\text{models}}} \sum_{j \in \{x,y,z\}} \left( F_{ij} - \bar{F}_j \right)^2 }
\end{equation}
where $\bar{F}_j$ is the mean of the force components over the committee.


\section{ASW characterization}\label{sec:ap_asw}

Reconstruction of the surface has been carried out using triangulation techniques, where the surface is approximated by a mesh of finite triangular elements.
A Triangular 3D Surface Plot (Tri-Surface Plot) is a type of surface plot constructed by triangulating compact surfaces using a finite number of triangles. Each point on the surface lies within one of these triangles. The intersection between any two triangles is either empty, a shared edge, or a shared vertex.
The triangulation took into account solely atoms that belongs to the surface of the periodic ice models, based on their Z value. The step size of the grid is 1.5 \text{\AA}.
All surface plots are created using \textit{ax.plot\_trisurf()}
 function of Matplotlib library\citep{hunter_matplotlib_2007}.
\begin{figure*}
    \centering
    \includegraphics[width=\linewidth]{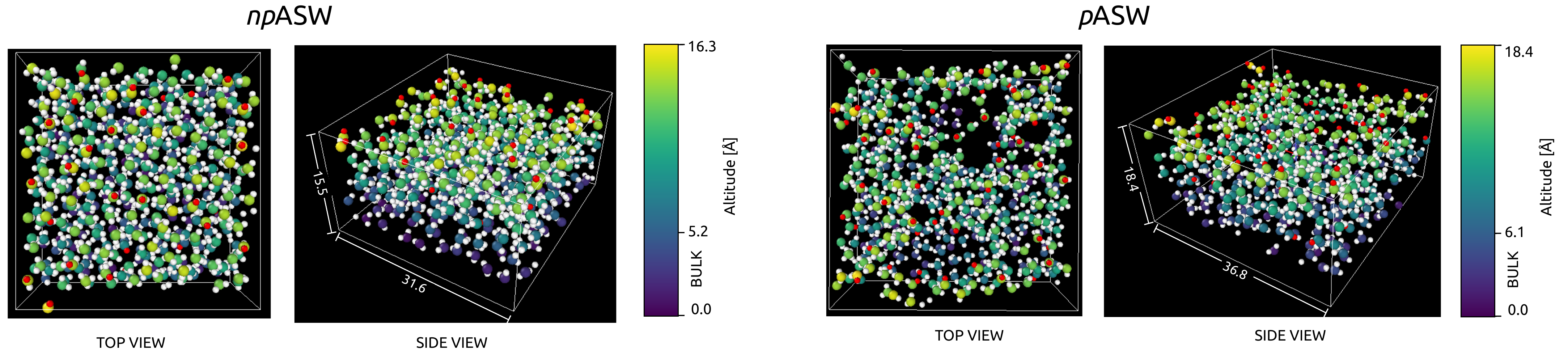} \\
    \caption{Left: Top and side view of one of the \textit{np}ASW periodic ice models. Cell dimensions are reported in \text{\AA}. Oxygen atoms are colored according to their altitude. Hydrogen atoms are colored in red if they are dangling-OH bond (i.e. non engaged in any HB-bond), else they are colored in  white. Right: analog for one of the \textit{p}ASW ice models. All atoms below the 5.2 or 6.1 \text{\AA} threshold (for \textit{np}ASW and \textit{p}ASW, respectively) are considered bulk and kept frozen at each stages of the work.}
    \label{fig:dOH_p}
\end{figure*}

The areal average roughness ($S_\text{a}$) of the ASW model is calculated as:
\begin{equation}
\label{eq:Ra}
S_\text{a} = \frac{1}{N} \sum_{i=1}^{N} \left| Z_i - \bar{Z} \right|
\end{equation}
where $Z_i$ are the individual height values, $\bar{Z}$ is the mean height of the surface and N is the number of points.
The Mean Roughness Depth ($R_z$) is calculated as the average of the five largest peak-to-valley differences in the surface:
\begin{equation}
\label{eq:rz}
    R_z = \frac{1}{5} \sum_{j=1}^{5} \left( Z_{\max, j} - Z_{\min, j} \right)
\end{equation}
where $Z_{max,j}$ and 
$Z_{min,j}$
are the highest and lowest points in each of the five largest peak-to-valley differences.
A dangling OH bond, d(OH), is a surface H atom not engaged in any H-bonds with other water molecules. In order to estimate the amount of d(OH) we assumed a threshold of 2.5 \text{\AA} in the shortest \ce{H_{W1}}--\ce{O_{W2}} distance for two different water molecules W1, W2.
The percentage of dangling OH bonds (\% $d$(OH)) is defined as: 
\begin{equation}
\label{eq:dOH}
    \%d(OH) = \frac{1}{N_\mathrm{models}} \sum_{j=1}^\mathrm{models} (\frac{N_{d(\mathrm{OH})}}{N_\mathrm{H}})100
\end{equation} 
For the 5 \textit{np}ASW and \textit{p}ASW surfaces;  where $N_{d(\mathrm{OH})}$ is the total number of dangling OH and $N_\mathrm{H}$ is the total number of surface H atoms, estimated excluding bulk H atoms.

\section{Gaussian bootstrap procedure}
\label{sec:ap_bootstrap}

To fit a Gaussian curve to the BE distribution taking into account the uncertainty of each binding energy derived from the three MLP models used in the query‐by‐committee approach, we employed a bootstrap method with 10000 simulations. In our analysis, each BE measurement, \(x_i\), is associated with an uncertainty \(\sigma_i\) that reflects the spread among the three model predictions. We assume that the true value of each measurement is normally distributed about \(x_i\) with variance \(\sigma_i^2\):
\begin{equation}
x_i^{\text{true}} \sim \mathcal{N}(x_i,\,\sigma_i^2).
\end{equation}
In each bootstrap simulation, we generate perturbed data by adding Gaussian noise to each measurement:
\begin{equation}
x_{i,\text{noisy}} = x_i + \epsilon_i,\quad \epsilon_i \sim \mathcal{N}(0,\,\sigma_i^2).
\end{equation}
For each simulation, we compute the sample mean and sample standard deviation:
\begin{equation}
\mu_{\text{sim}} = \frac{1}{N_{\text{data}}} \sum_{i=1}^{N_{\text{data}}} x_{i,\text{noisy}},
\end{equation}
\begin{equation}
\sigma_{\text{sim}} = \sqrt{\frac{1}{N_{\text{data}}-1} \sum_{i=1}^{N_{\text{data}}} \left(x_{i,\text{noisy}} - \mu_{\text{sim}}\right)^2},
\end{equation}
where \(N_{\text{data}}\) is the number of measurements. This procedure is repeated \(N=10000\) times. The final estimates for the mean and standard deviation are then obtained as the averages over the simulations:
\begin{equation}
\mu_{\text{estimate}} = \langle \mu_{\text{sim}} \rangle
\qquad
\sigma_{\text{estimate}} = \langle \sigma_{\text{sim}} \rangle.
\end{equation}
The uncertainties in these estimates are given by the standard deviations of the corresponding bootstrap distributions:
\begin{equation}
\sigma_{\mu} = \sqrt{\langle (\mu_{\text{sim}} - \mu_{\text{estimate}})^2 \rangle}
\quad
\sigma_{\sigma} = \sqrt{\langle (\sigma_{\text{sim}} - \sigma_{\text{estimate}})^2 \rangle}.
\end{equation}
The Gaussian probability density function is defined as
\begin{equation}
f(x) = \frac{1}{\sqrt{2\pi}\,\sigma_{\text{estimate}}} \exp\left(-\frac{(x - \mu_{\text{estimate}})^2}{2\sigma_{\text{estimate}}^2}\right).
\end{equation}

To quantify the uncertainty in the histogram of the BE distribution, we compute error bars for each histogram bin based on the bootstrap simulations. Let \(n_{i,j}\) denote the count in the \(i\)-th bin for the \(j\)-th bootstrap simulation, with \(j=1,\dots,N\). The mean count in bin \(i\) is given by
\begin{equation}
\langle n_i \rangle = \frac{1}{N} \sum_{j=1}^{N} n_{i,j},
\end{equation}
where N = 10000 and the error bar for bin \(i\), representing the uncertainty in the count due to the propagated measurement errors, is calculated as
\begin{equation}
\sigma_{n_i} = \sqrt{\frac{1}{N-1} \sum_{j=1}^{N} \left(n_{i,j} - \langle n_i \rangle\right)^2}.
\end{equation}
This approach propagates the individual uncertainties from the neural network models into the Gaussian fitting procedure.

\section{Interaction energy incremental analysis}\label{sec:ap_sapt}

\begin{figure*}
    \centering
    \includegraphics[width=0.8\linewidth]{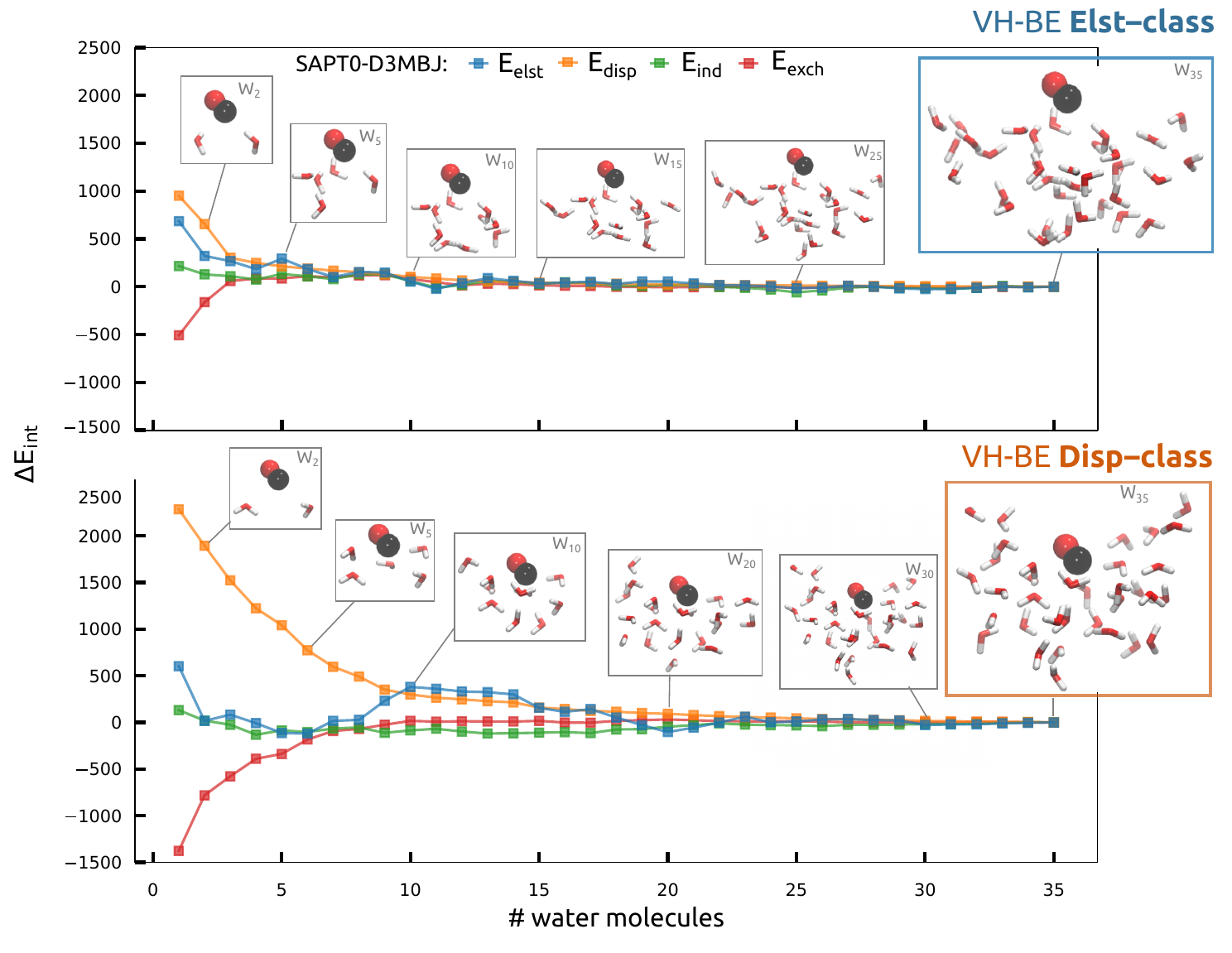}
    \caption{Convergence of SAPT0-D3MBJ interaction energy components for CO as a function of binding site  size. The plot is generated by incrementally adding water molecules to CO from an extracted VH-BE Elst-class binding site (upper panel) and a VH-BE Disp-class binding site (lower panel). The interaction energy difference are with respect of the CO + W$_{35}$ limit ($\Delta E_{int} = E_{int}(CO+W_{35}) - E_{int}(CO+W_{X}$). The interaction energy in the Elst-class converges significantly faster than in the Disp-class binding site.
    The interaction energy is further decomposed into electrostatic, induction, dispersion and exchange contribution, of which the dispersion interaction is the 
    slowest to converge. Therefore, selecting a binding site  extract  comprising 28 water molecules captures almost the totality of the interaction energy of CO + ASW.}
    
    \label{fig:sapt_incr}
\end{figure*}

We examined how the different contributors to the interaction energy converges in relation 
to the size of the binding sites, for two VH-BE example structures belonging to different classes, also reported in Fig. \ref{fig:example_classes}. Fig. \ref{fig:sapt_incr} illustrates the results  
as the binding site expands from 2 to 35 water molecules around the CO adsorption location, in the stationary-point geometry. The 0 K baseline corresponds to the original energy value for the CO\ce{-W_{35}} structures, such that the plotted energies represent the deviation at each cluster size.
For both structures, the largest attractive contribution to the interaction energy comes from the dispersion, followed by electrostatic and induction energy, while the only repulsive interaction corresponds to the exchange energy.
 
 For the Elst-class structure, upper panel, all the contributes are converged within the first 10 water molecules. This confirms the rather local character of the primarily electrostatic interaction  between CO and the 
ASW surface, as recovering the full interaction
energy requires only a small portion of the water molecules near the binding 
site.  
On the other hand, for the Disp-class structure, lower panel, the dispersion energy contribution is the slowest to converge, as expected, 
and reaches full convergence when the binding site has grown to $\sim$ 20 water molecules.
Moreover, al contributions to the interaction energy are larger than  the El-structure, which is consistent with a CO molecule further embedded into the ASW H-bond network.

\begin{figure}
    \centering
    \includegraphics[width=0.5\linewidth]{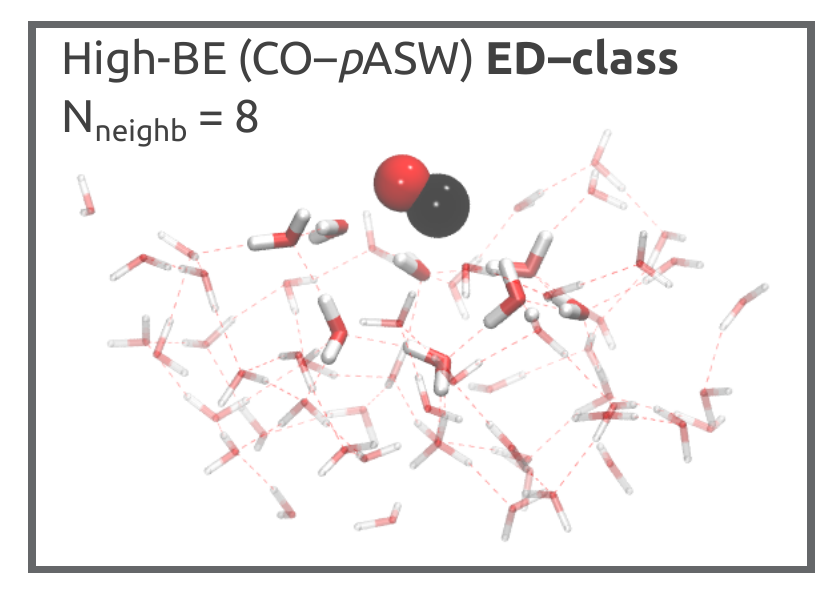} 
    \caption{Example of Very High (VH)-BE structure belonging to ED-class (BE = 1630 K). The figure  displays a portion of the binding site comprising 50 water molecules, represented as sticks. The water molecules within 4.5 $\text{\AA}$ of CO center of mass (i.e. nearest neighbors) have been highlighted.}
\label{fig:ed_class}
\end{figure}

\section{Simulated TPD curve generation}\label{sec:ap_TPD}

The simulated temperature-programmed desorption (TPD) traces were generated based on a first-order desorption model. 

Desorption rates for each population were computed using a first-order Arrhenius expression:

\begin{equation}
k_{\mathrm{des},i}(T) = \nu \exp\left(-\frac{E_i}{k_\text{B}T}\right),
\end{equation}
with \(\nu\) as the pre-exponential factor (s\(^{-1}\)).

In order to simulate the temperature evolution during a TPD experiment, 
we used a linear heating ramp:

\begin{equation}
T(t) = T_0 + \beta t,
\end{equation}

with \(T_0\) as the initial temperature and \(\beta\) as the heating rate (K/s). Surface coverage \(\theta_i(t)\) for each population was obtained by integrating the differential equation:

\begin{equation}
\frac{d\theta_i}{dt} = -\nu \exp\left(-\frac{E_i}{T(t)}\right) \theta_i.
\end{equation}

In order to consider the complete BE distribution, it was discretized into 40 bins.
For each bin, the bin center is taken as the representative binding energy, \(E_i\), 
and the weight for that bin is computed as follows. If \(n_i\) is the number of states 
falling into the \(i\)th bin, then the weight is defined as
\begin{equation}
w_i = \frac{n_i}{\sum_{j=1}^{N_\mathrm{bins}} n_j},
\end{equation}
so that the weights satisfy
\begin{equation}
\sum_{i=1}^{N_\mathrm{bins}} w_i = 1.
\end{equation}

On the other hand, the instantaneous desorption flux from each population is given by: \\
\begin{equation}
    F_i(T) = \cdot \left(-\beta \cdot \frac{d\theta_i}{dt}\right),
\end{equation}

The individual fluxes are summed to yield the overall TPD signal:

\begin{equation}
F_{\mathrm{total}}(t)=\sum_{i=1}^{N_\mathrm{bins}} w_i F_i(t).
\end{equation}

Furthermore, in TPD experiments, CO molecules initially bound to low BE sites often diffuse to higher BE 
sites before desorption. To more accurately reflect this phenomenon, we applied minimum BE cutoffs.  
This approach considers only desorption events originating from sites with BE values 
exceeding the cutoff, effectively mimicking the observed migration and subsequent desorption 
from higher BE sites in real TPD scenarios
Hence, for each cutoff, only sites with \(E \geq E_{\mathrm{min}}\) were retained. 
To consider the lowering in  surface coverage, initial 
coverage \(\theta_0\) was scaled by the fraction \(f\) of states surviving the cutoff:

\begin{equation}
f = \frac{N_{\mathrm{truncated}}}{N_{\mathrm{all}}}, \quad \theta_0 = f,
\end{equation}

where \(N_{\mathrm{filtered}}\) is the number of states remaining after the cutoff, and \(N_{\mathrm{all}}\) is the total number of states. Increasing the cutoff reduces the effective 
initial coverage, resulting in desorption peaks shifting to higher temperatures and lower 
integrated TPD signals. This approach effectively captures the desorption dynamics arising 
from discrete adsorption populations and surface diffusion effects.

\end{appendix}

\end{document}